\documentclass[a4paper,11pt]{article}
\pdfoutput=1 

\usepackage{jinstpub} 
\title{\boldmath Results on Multiple Coulomb Scattering from 12 and 20 GeV electrons on Carbon targets}

\author{G. Abbiendi$^a$,
J. Bernhard$^b$,
F. Betti$^{a,c}$,
M. Bonanomi$^d$,
C. M. Carloni Calame$^e$,
M. Garattini$^{b,g}$,
Y. Gavrikov$^f$,
G. Hall$^g$,
F. Iacoangeli$^h$,
F. Ignatov$^i$,
M. Incagli$^j$,
V. Ivanchenko$^{b,k}$,
F. Ligabue$^{j,l}$,
T. O. James$^g$,
U. Marconi$^a$,
C. Matteuzzi$^d$,
M. Passera$^m$,
M. Pesaresi$^g$,
F. Piccinini$^e$,
R. N. Pilato$^{j,n}$,
F. Pisani$^{a,b,c}$,
A. Principe$^{a,c}$,
W. Scandale$^b$,
R. Tenchini$^j$, and
G. Venanzoni$^{j,1}$\note{Corresponding author}}

\affiliation[a]{INFN Sezione di Bologna, Viale Carlo Berti-Pichat 6/2, 40127 Bologna, Italy}
\affiliation[b]{CERN, 1211 Geneva 23, Switzerland}
\affiliation[c]{Universit\`a di Bologna, Bologna, Italy}
\affiliation[d]{INFN Sezione di Milano-Bicocca e Universit\`a Milano-Bicocca, Piazza della Scienza 3, 20126 Milano}
\affiliation[e]{INFN Sezione di Pavia, Via Agostino Bassi 6, 27100 Pavia, Italy}
\affiliation[f]{Petersburg Nuclear Physics Institute in National Research Centre "Kurchatov Institute", 188300 Gatchina, Russia}

\affiliation[g]{Blackett Laboratory, Imperial College London, London SW7 2AZ, U.K}
\affiliation[h]{INFN Sezione di Roma, Piazzale Aldo Moro 2, 00185 Rome, Italy}
\affiliation[i]{Budker Institute of Nuclear Physics,  SB RAS, and Novosibirsk State University, Novosibirsk, 630090, Russia}
\affiliation[j]{INFN Sezione di Pisa, Largo Bruno Pontecorvo 3, 56127 Pisa, Italy}
\affiliation[k]{Tomsk State University, 634050 Tomsk, Russia}
\affiliation[l]{Scuola Normale Superiore, Pisa, Italy}

\affiliation[m]{INFN Sezione di Padova, Via Francesco Marzolo 8, 35131 Padua, Italy}
\affiliation[n]{Dipartimento di Fisica, Universit\`a di Pisa, I-56127 Pisa, Italy}

\emailAdd{graziano.venanzoni@pi.infn.it}

\abstract{Multiple scattering effects of 12 and 20 GeV electrons on 8 and 20 mm thickness carbon targets have been studied with high-resolution silicon microstrip detectors of the UA9 apparatus at the H8 line at CERN.
  Comparison of the scattering angle between data and GEANT4 simulation shows excellent agreement in the core of the distributions leaving some residual disagreement in the tails.}


\begin{document}
\maketitle
\flushbottom

\section{Introduction}
\label{sec:intro}
A precise knowledge of the multiple Coulomb scattering  is required in many experimental and theoretical activities. Recently a new experiment, MUonE, has been proposed with the aim of measuring the running of the effective electromagnetic coupling at low momentum transfer in the space-like region ($\alpha(q^2), q^2<0$) to provide an independent determination of the leading hadronic contribution to the muon g-2~\cite{Calame:2015fva,Abbiendi:2016xup}.  Such a measurement relies on the precise determination of the measured angle of the electrons elastically scattered from high-energy muons (150 GeV) impinging on 1-2 cm beryllium targets~\cite{Abbiendi:2016xup}.
Multiple scattering effects of electrons within the target limit the accuracy of the reconstructed angle and must be kept under control at percent level~\cite{graziano}.
Such an accuracy requires a precise tuning of the Monte Carlo (MC) simulation on real data.
In the following we discuss the results of a beam test performed on the H8 line at CERN with electrons of 12 and 20 GeV on carbon targets of different thickness and we compare them with the predictions from a Monte Carlo simulation based on GEANT4~\cite{Agostinelli:2002hh,Allison:2006ve,Allison:2016lfl}. 
The experimental set-up employed for this beam test is commonly used by the UA9 collaboration~\cite{Scandale:2011mq} to study crystal channeling with high-energy proton and pion beams. The measurement greatly benefits from the precise tracking of the  high-resolution silicon microstrip detectors of the UA9 apparatus~\cite{Pesaresi:2011zz}. \\

\section{Beam test setup}
The used set-up consists of two upstream planes of Si trackers (planes 1 and 2) separated by a distance of $ \sim $10 m, a target and three downstream tracking planes (planes 3, 4, and 5) with a lever arm of $ \sim $ 1 m. Each silicon tracker is composed of two layers 320 $ \mu$m thick, with a $ 3.8\times 3.8\,\text{cm}^2 $ active area, to measure both the x and y coordinates (in the plane orthogonal to the beam, taken as z axis), with an intrinsic hit resolution of $7~\mu$m, resulting in 0.02 mrad angular resolution of the telescope.
Events were triggered on the coincidence of signals from scintillator slabs placed upstream of the telescope.
Two targets of isostatic graphite (1.83 g/cm$^{3}$ density, 23.32 cm radiation length) of 8 and 20 mm thickness were tested.

\begin{figure}
	\centering
        \includegraphics[width=12cm]{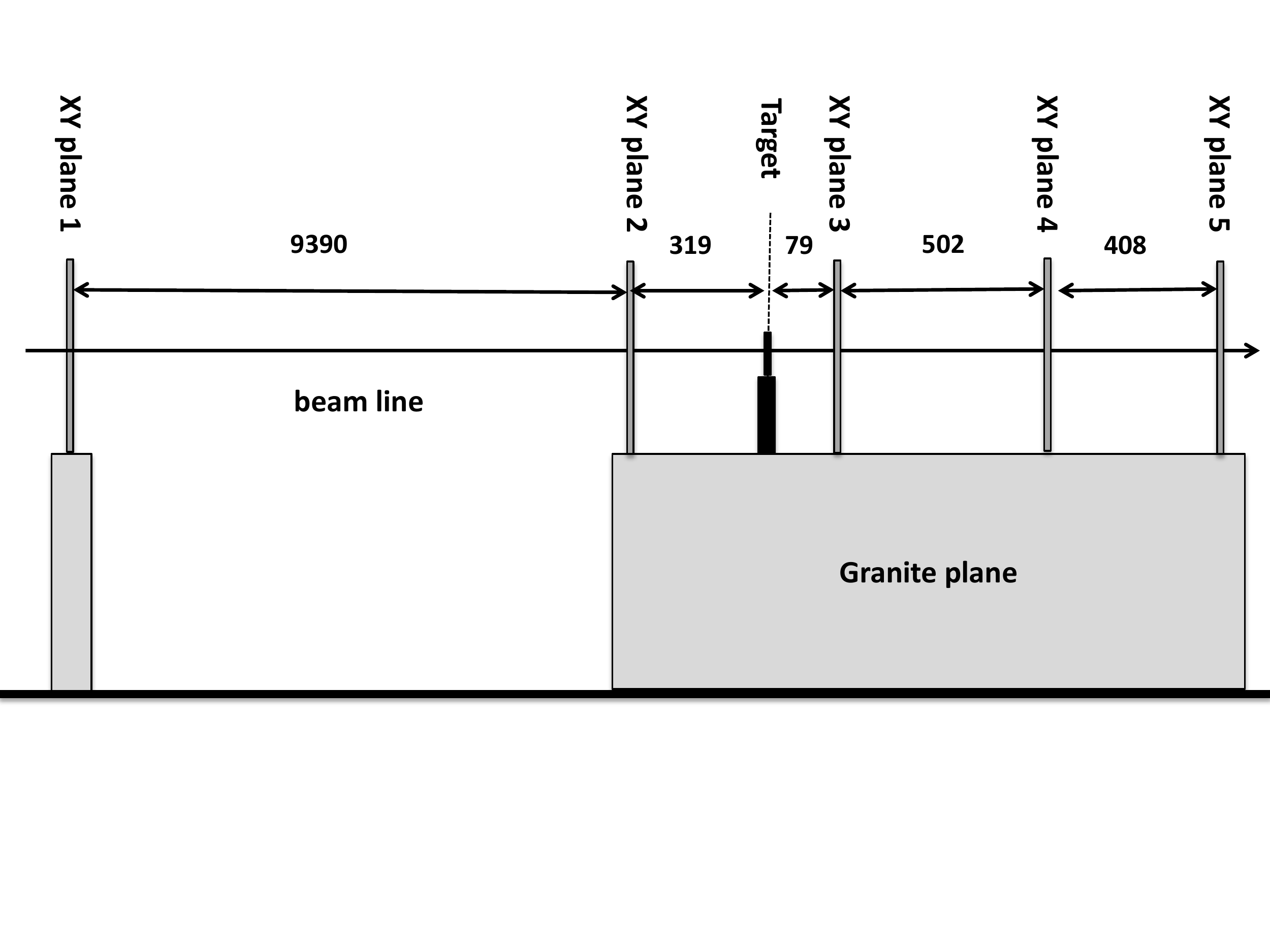}
	\caption{Beam test detector set-up. Distances are given in mm. The beam direction defines the z axis.}
	\label{img:setup17}
\end{figure}

\begin{table}[htbp]
	\centering
	\begin{tabular}{ccc}
\hline
	  \textbf{Beam} & \textbf{Target Type} & \textbf{N events}$ \times $10$ ^6 $\\
\hline
		12 GeV e$ ^- $ & 8 mm C & 15\\
		20 GeV e$ ^- $ & 8 mm C & 12\\
		12 GeV e$ ^- $ & 20 mm C & 15\\
\hline\hline
	\end{tabular}
	\caption{Data runs with target.}
	\label{tab:normal_runs}
\end{table}
\begin{table}[htbp]
  \centering
	\begin{tabular}{lcc}
\hline
          	\textbf{Beam}  & \textbf{N events}$ \times $10$ ^6 $\\ 
\hline
		80 GeV $ \pi^+ $ & 8\\
		20 GeV e$ ^- $ & 1 \\
		12 GeV e$ ^- $ & 1\\
		160 GeV $ \mu^+ $ & 2 \\
		180 GeV $ \pi^+ $ & 5\\
\hline\hline
	\end{tabular}
	\caption{Alignment runs (no target).}
	\label{tab:align_runs}
\end{table}

Figure \ref{img:setup17} shows the beam test set-up with the distances given in mm. 
Data were taken with electron momenta of 12 and 20 GeV/c, with different target thickness, as reported in Table~\ref{tab:normal_runs}.
For a determination of the contribution of the telescope, data without the carbon target (alignment runs) were taken, as shown in Table~\ref{tab:align_runs}.

\section{Alignment}
\label{FidCut}
\begin{figure}[tb]
	\centering
	 \includegraphics[scale=0.6]{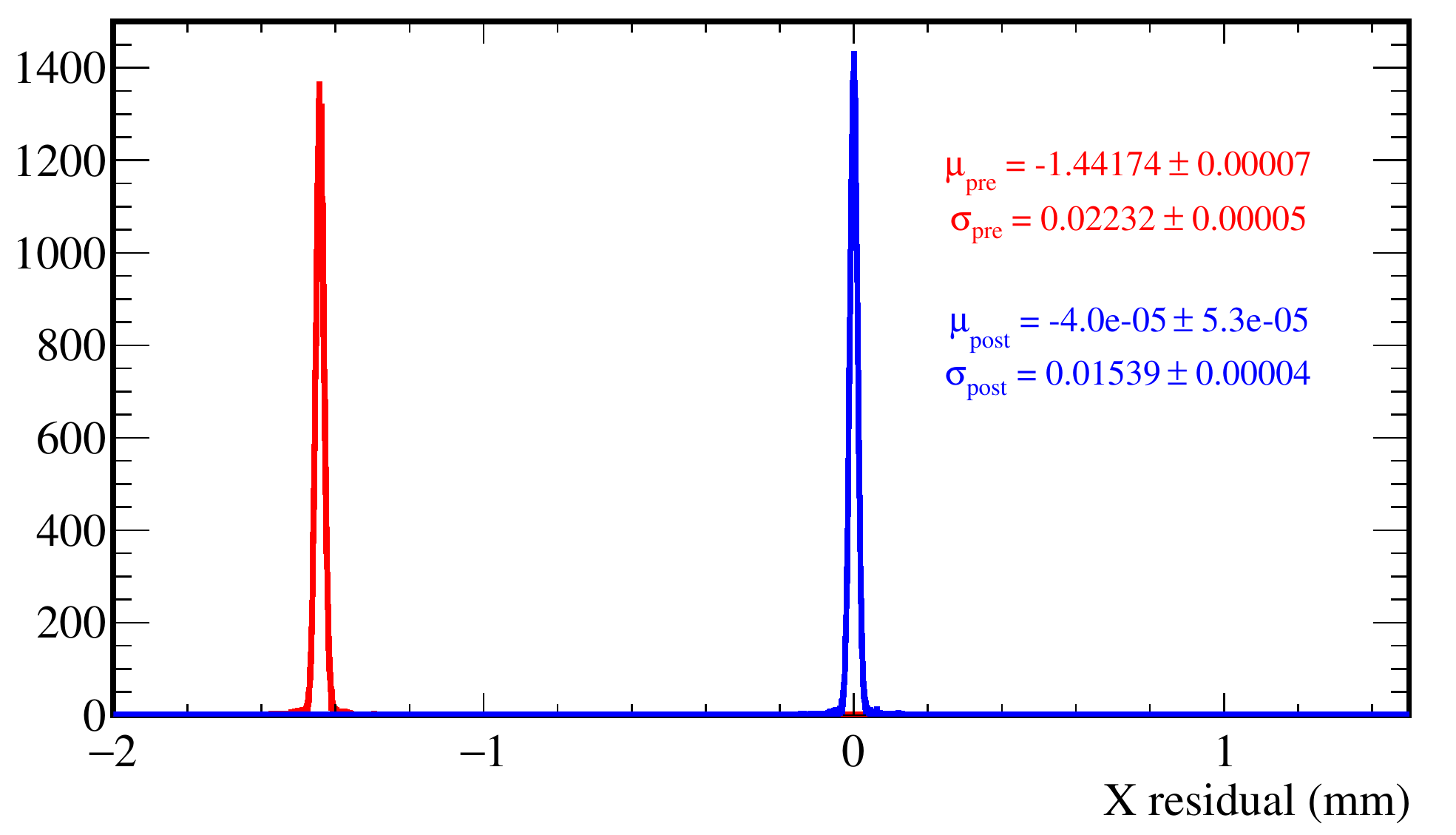}
	\caption{Residual before (red histogram) and after (blue histogram) the alignment.}
	\label{fig:align_test}
\end{figure}

The alignment procedure is based on a recursive algorithm minimizing track residuals.
Pions at 80 and 180~GeV (without target) have been used to determine the alignment parameters.
Only tracks that have generated a single hit in each sensor are selected and 
straight line tracks have been defined through the selected planes (1 and 2) in the upstream region.
Straight lines are then extrapolated to the downstream region (planes 3, 4, 5) to get the expected coordinates and measure hit residuals (defined as differences between measured and estimated positions).
Hit residuals are minimized in both views and rotational misalignments (which appear as a correlation between the hit positions in one view and residuals in the other)
corrected by an iterative procedure~\cite{Pesaresi:2011zz}.
Figure~\ref{fig:align_test} shows an example of the  mean value of the residual distributions before and after the alignment procedure.

The high statistics of the data samples used for the alignment allows to reach an accuracy in the correction of the transverse offsets of 1~$\mu$m, and 
less than 0.1~mrad for the rotational angles (x-y plane), as quoted in ref.~\cite{Pesaresi:2011zz}.
Residual misalignment in the longitudinal direction, which is more difficult to correct with data, is estimated to be of the order of 1~mm.

\section{Event selection}
Once the alignment parameters have been determined, they are used
in the electron runs.
The telescope measures an incoming ($\theta_{IN}$) and outgoing ($\theta_{OUT}$)  track angle at the target where $\theta_{IN}$  and $\theta_{OUT}$ are the angles with respect to the $z$ axis obtained from upstream and downstream tracking planes respectively.
The projection in the x (y) coordinate of the scattering angle $\Delta\theta_X$ ($\Delta\theta_Y$) is defined as $\Delta\theta_X  = \theta_{OUT,X}-\theta_{IN,X}$ ($\Delta\theta_Y = \theta_{OUT,Y}-\theta_{IN,Y}$) where  $\theta_{OUT,X}$ and $\theta_{IN,X}$ ($\theta_{OUT,Y}$ and $\theta_{IN,Y}$) are the $x$ ($y$) projections of the outcoming and incoming angles.

\begin{figure}[ht]
   \includegraphics[width=.5\textwidth]{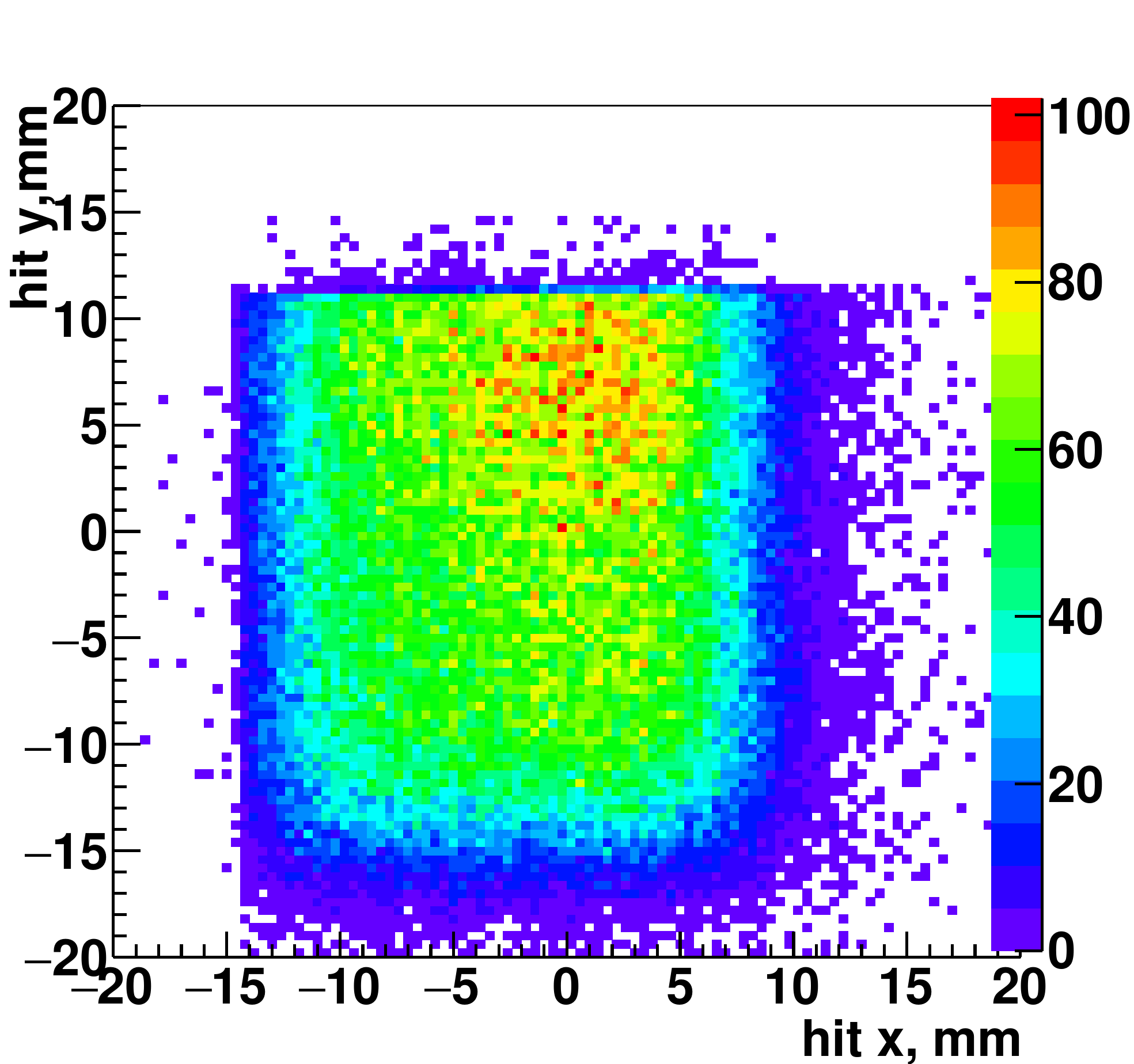}~\includegraphics[width=.5\textwidth]{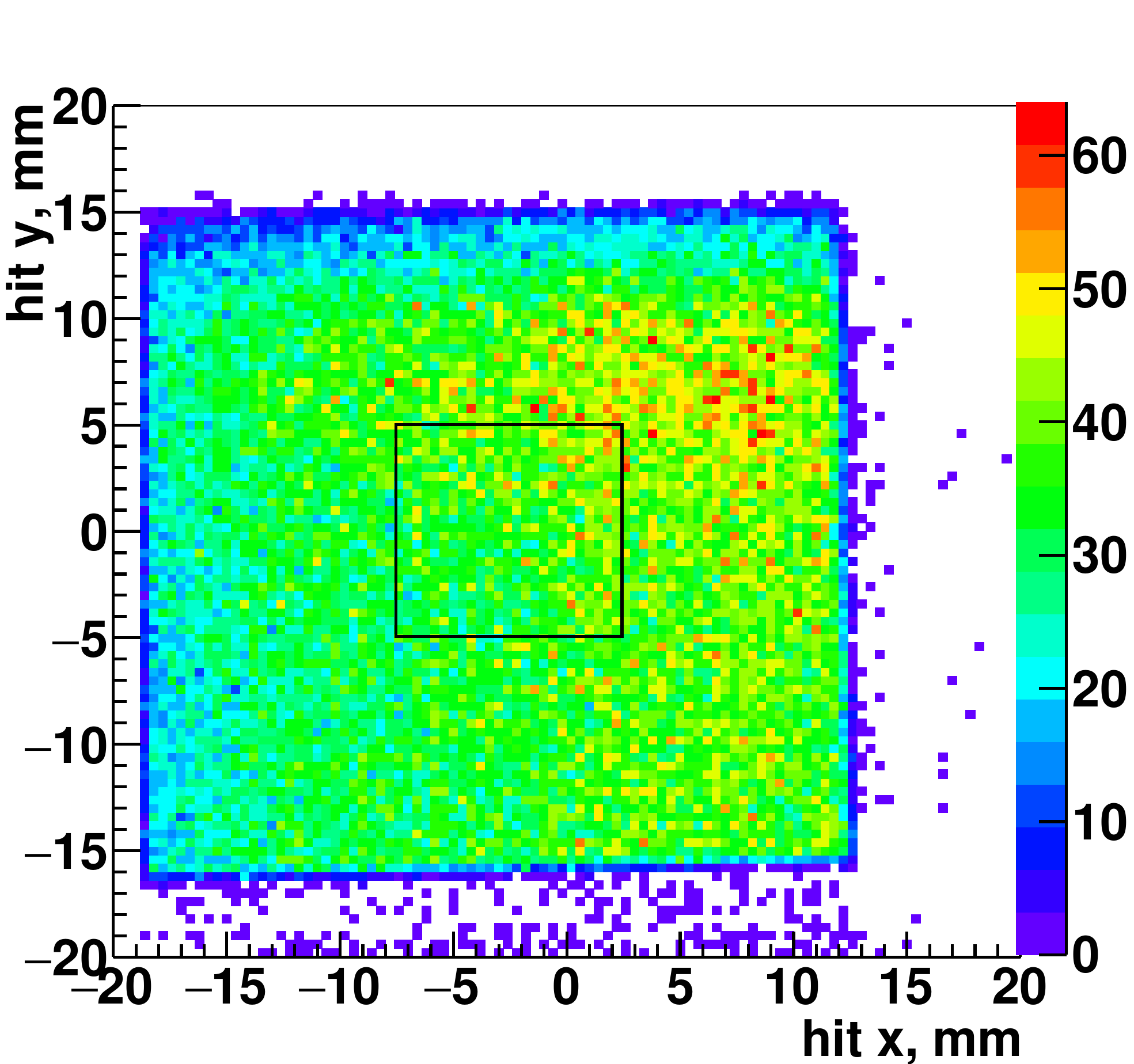}
 
\caption{Hit occupancy of the electron beam of 12 GeV in the upstream region, (Left) first plane; (Right) second plane.}
\label{img:xprof1}
\end{figure}

Differently from pions and muons, the beam spot of electrons~\footnote{For pions and muons the beam spot is a few mm wide.} has a standard deviation of $\sim 3$ cm in both horizontal and vertical directions covering almost all the sensitive sensor area, as shown in Fig.~\ref{img:xprof1}. To ensure good quality events and full acceptance in the downstream region only single tracks are selected.
Track-associated hits inside a fiducial region of  $10\times10$ mm$^2$ in plane 2 (as shown by the box in Fig.\ref{img:xprof1}) together with $\theta_{IN}< 1$ mrad are required. About 1/10 of the data sample survives the selection cuts.

\begin{figure}
  \includegraphics[width=.5\textwidth]{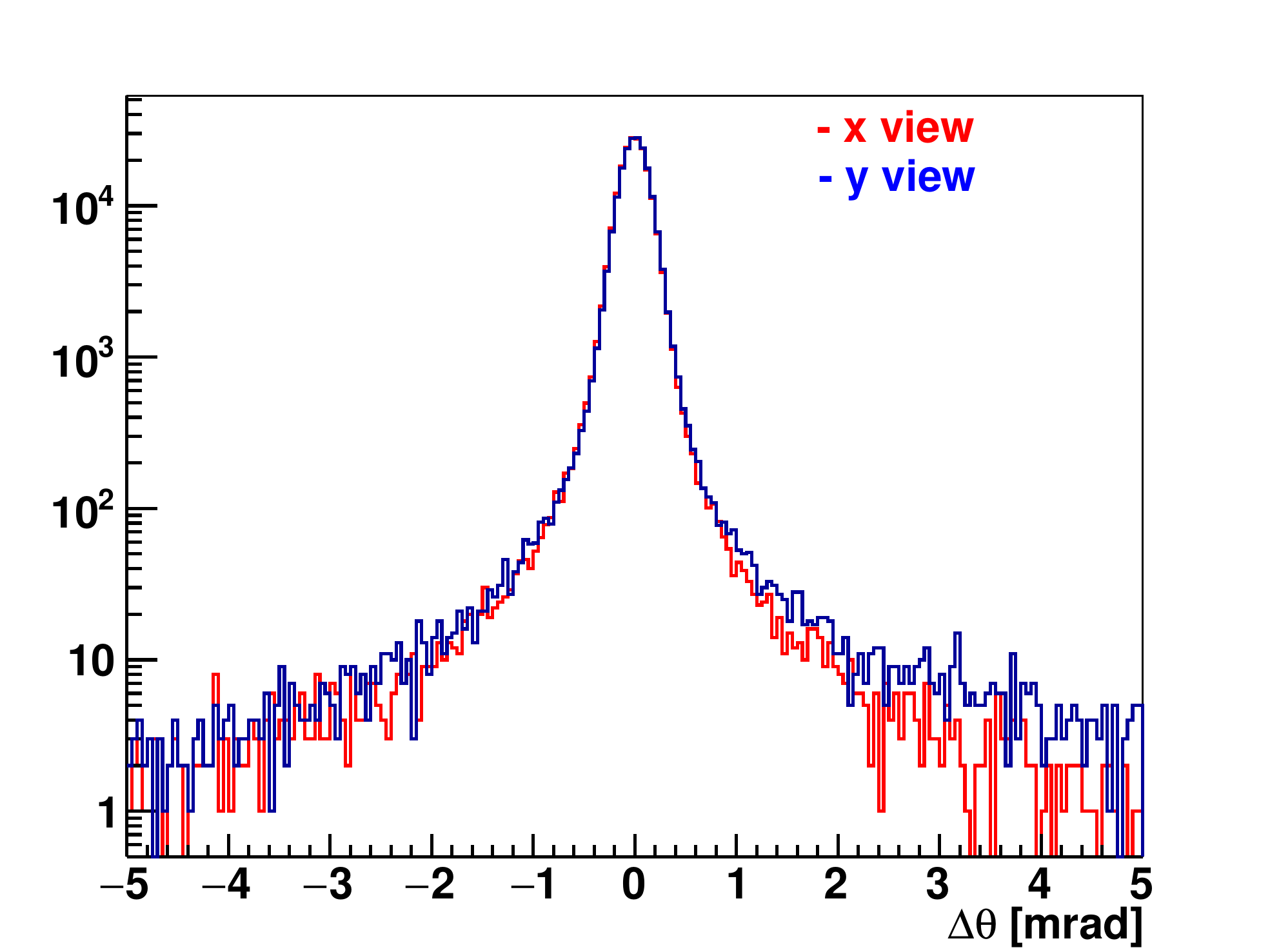}~\includegraphics[width=.5\textwidth]{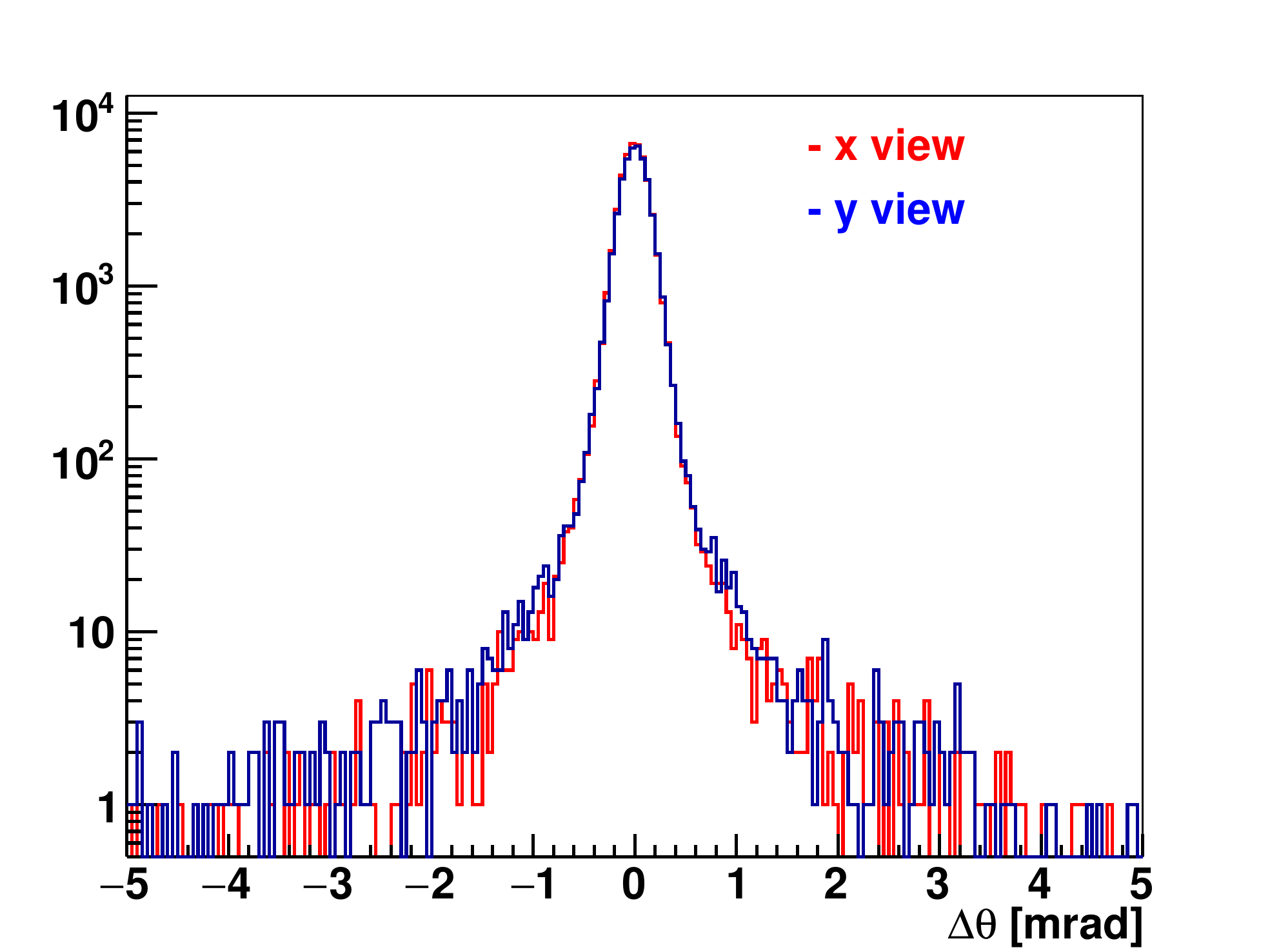} 
  \caption{Angular deflections in both views: (Left) without and (Right) with fiducial cuts.
}
	\label{img:12GeV_cut}
\end{figure}

Figure \ref{img:12GeV_cut} shows the effect of these cuts on the distribution of the scattering angle of 12 GeV electrons without target. As can be seen, these cuts remove most of the tails and make the angular distribution more symmetric.
In particular by defining the asymmetry as the number of events with $\Delta\theta_{X,Y}<-1$ mrad minus the ones with $\Delta\theta_{X,Y}>1$ mrad divided by their sum, after applying the fiducial cuts the asymmetry is consistent with zero.
The result presented here for the 12 GeV case holds for the 20 GeV data as well.

\section{Results}
In order to compare the multiple scattering on the target with the one expected from the GEANT4 simulation, we describe the $x$ and $y$ projection of the reconstructed scattering angle (indicated here with $\theta$) by a convolution of the detector and target effects:
\begin{equation}
  f(\theta)=f_{\text{telescope}}(\theta)\otimes f_{\text{target}}(\theta)
  \label{eq_conv}
\end{equation}
The contribution of the experimental set-up  ($f_{\text{telescope}}(\theta)$) was studied with alignment runs.
Corrections  to Eq.~\ref{eq_conv} due to energy loss and multiple scattering in the detector, which affect the energy and angular distribution of the $e^-$ on target,
were taken into account by simulation.
Particularly, the material budget in the upstream region reduces by about 5\% the energy of the electrons that reach the target.


\begin{figure}[h]
	\centering
        \includegraphics[width=12cm]{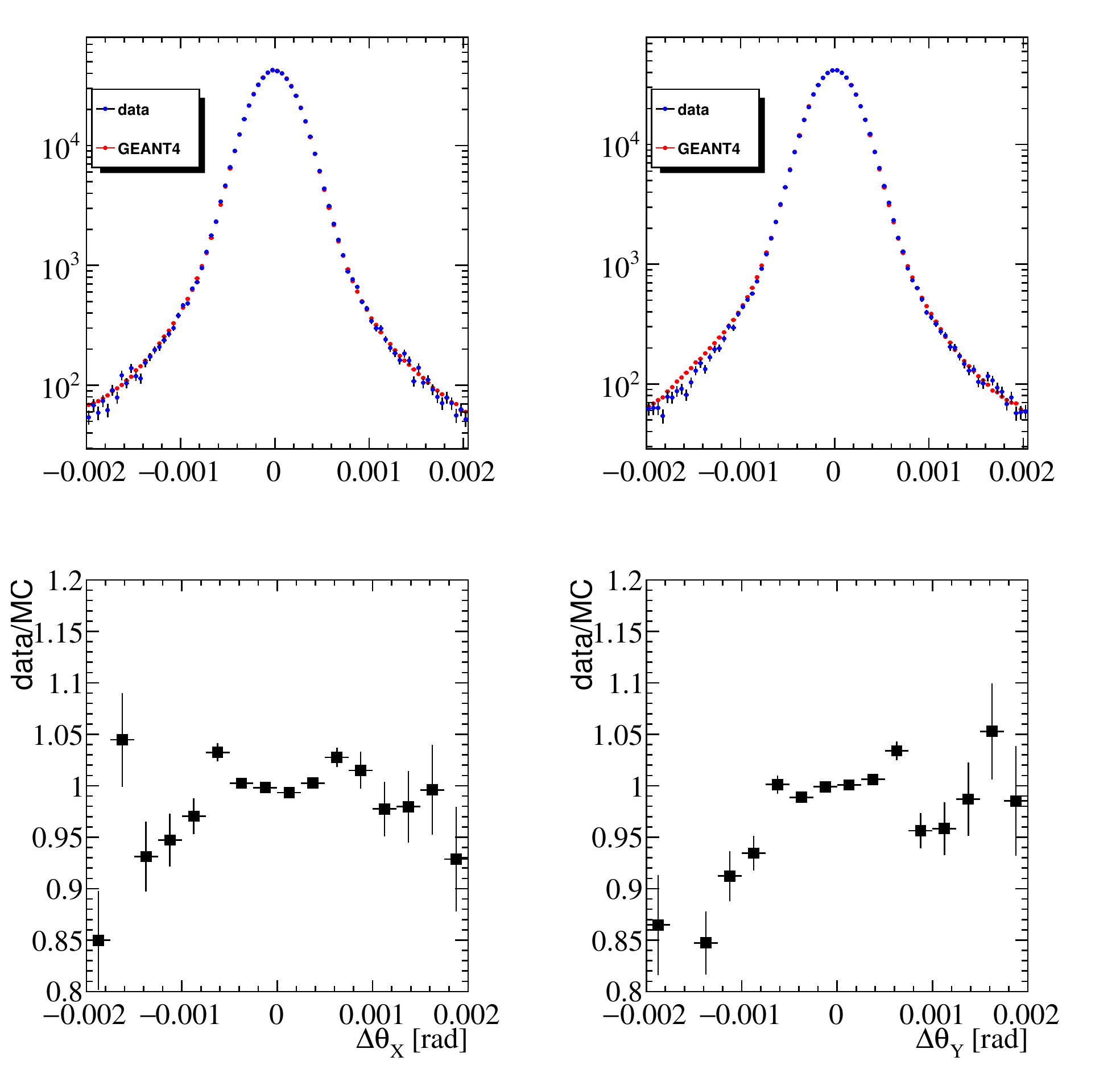}
	\caption{Left: (upper) $x$-projection of the scattering angle from data and GEANT4 for 12 GeV e$^-$ with 8 mm C target; (lower) data/MC ratio; Right: (upper)
          $y$-projection of the scattering angle from data and GEANT4 for 12 GeV e$^-$ with 8 mm C target; (lower) data/MC.}        
	\label{img:acc_cut1}
\end{figure}
\begin{figure}[h]
	\centering
        \includegraphics[width=12cm]{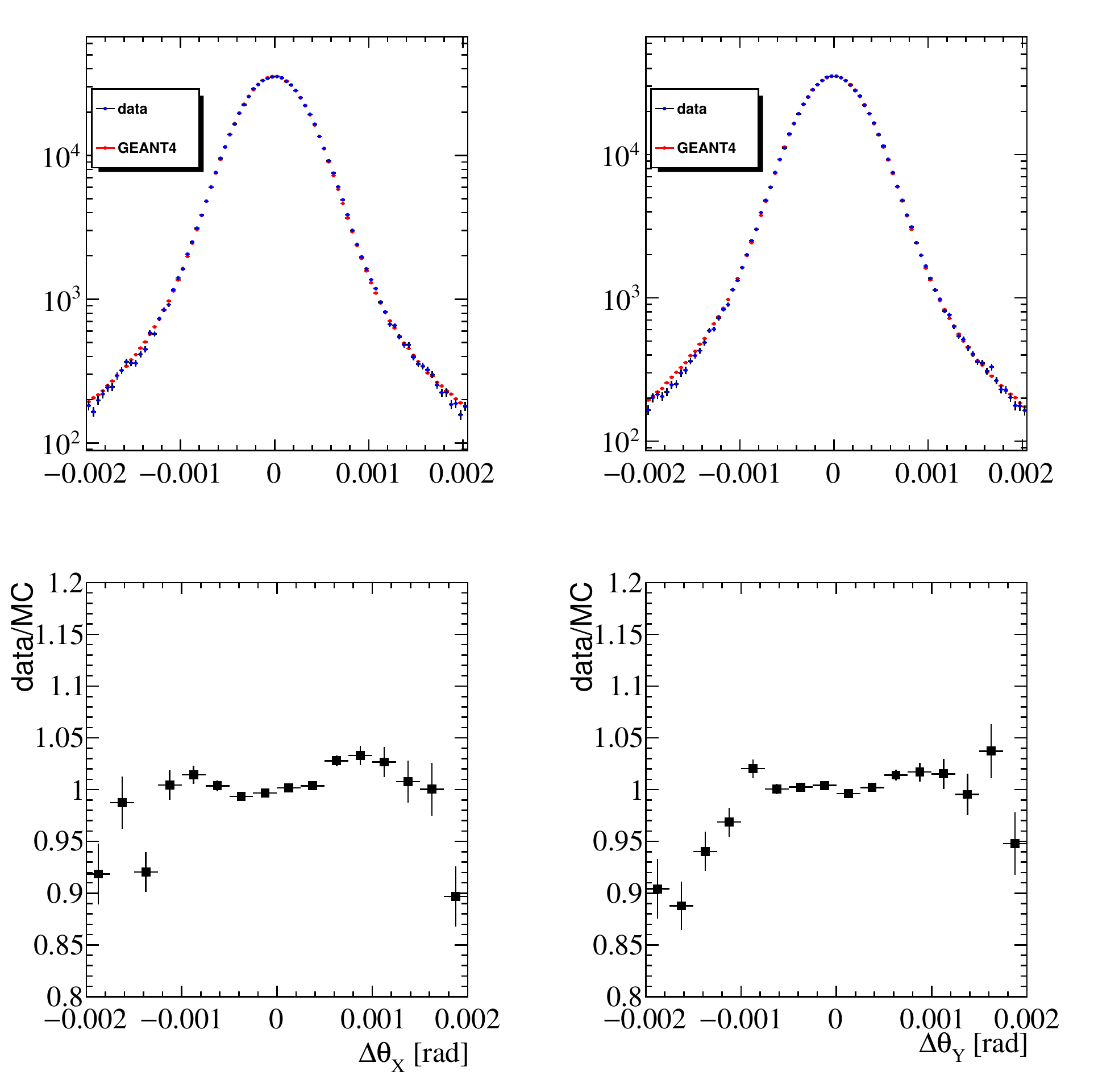}
	\caption{Left: (upper) $x$-projection of the scattering angle from data and GEANT4 for 12 GeV e$^-$ with 20 mm C target; (lower)  data/MC ratio; Right: (upper)
          $y$-projection of the scattering angle from data and GEANT4 for 12 GeV e$^-$ with 20 mm C target; (lower) data/MC ratio.}        
	\label{img:acc_cut2}
\end{figure}
\begin{figure}[h]
	\centering
        \includegraphics[width=12cm]{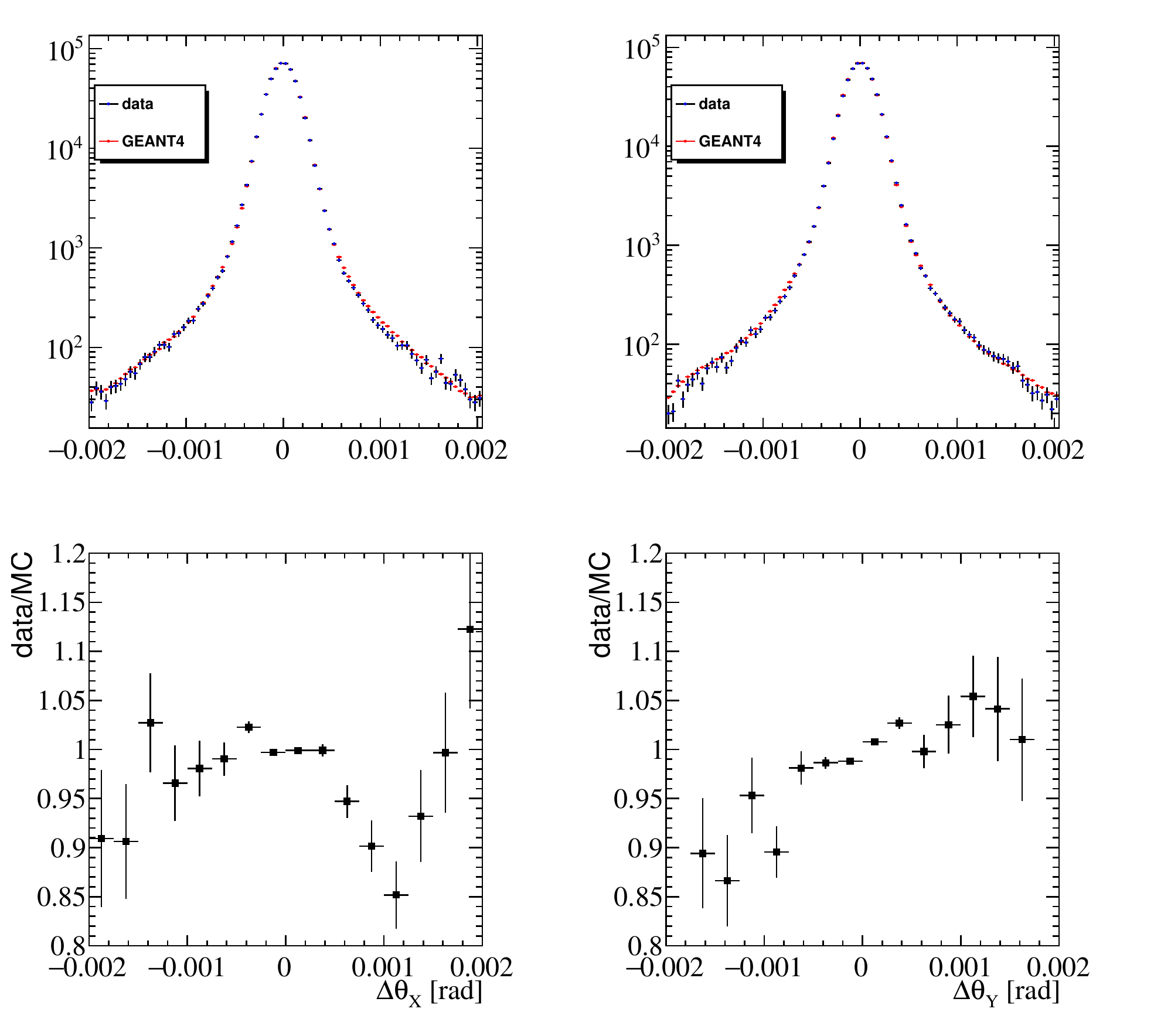}
	\caption{Left: (upper) $x$-projection of the scattering angle from data and GEANT4 for 20 GeV e$^-$ with 8 mm C target; (lower) data/MC ratio; Right: (upper)
          $y$-projection of the scattering angle from data and GEANT4 for 20 GeV e$^-$ with 8 mm C target; (lower) data/MC ratio.}        
	\label{img:acc_cut3}
\end{figure}

To simulate the target ($f_{\text{target}}(\theta)$ in Eq.~\ref{eq_conv}) version \texttt{10.4p02} of GEANT4 with  \texttt{Opt4} electromagnetic physics~\cite{geant} has been used, with the following multiple scattering model: \\
    \begin{verbatim}     
      GoudsmitSaunderson :  Emin=    0 eV    Emax=    100 MeV
      Table with 120 bins Emin=    100 eV    Emax=    100 MeV
      WentzelVIUni :      Emin=    100 MeV   Emax=     10 TeV
      Table with 100 bins Emin=    100 MeV   Emax=     10 TeV
\end{verbatim}
    Figures~\ref{img:acc_cut1}-\ref{img:acc_cut3}~show the comparison of the $x$ and $y$ projection of the scattering angle from data with the expected one from GEANT4 simulation. Distributions are normalized to the same number of events.
    As can be seen the overall agreement is quite satisfactory (at percent level), apart from the tail region, where differences reach 10\%, although with large statistical errors.
In the tail fraction of the distribution, differences between data and Monte Carlo can be due to residual misalignment or acceptance cuts which are not properly taken into account in the simulation (due to missing hit digitization) and to the limited statistics of the alignment data which were used to obtain the Monte Carlo distribution of the scattering angle (see Eq. 5.1). The effects from the apparatus which do not scale with the energy of the particle are expected to have a larger impact for 20 GeV electrons, where the multiple scattering effects are smaller.

\begin{figure}
	\centering
        \includegraphics[width=0.5\textwidth]{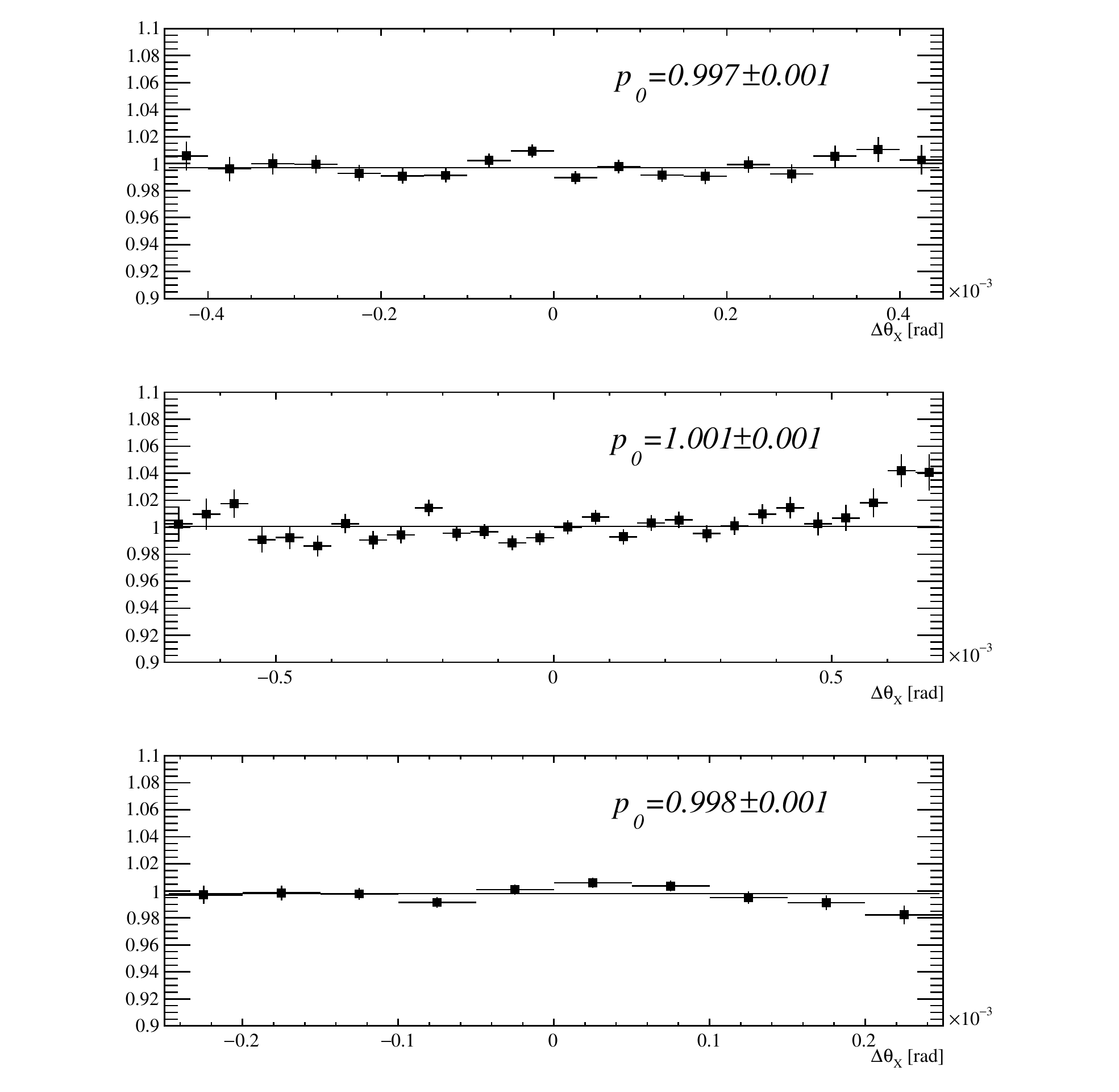}~\includegraphics[width=0.5\textwidth]{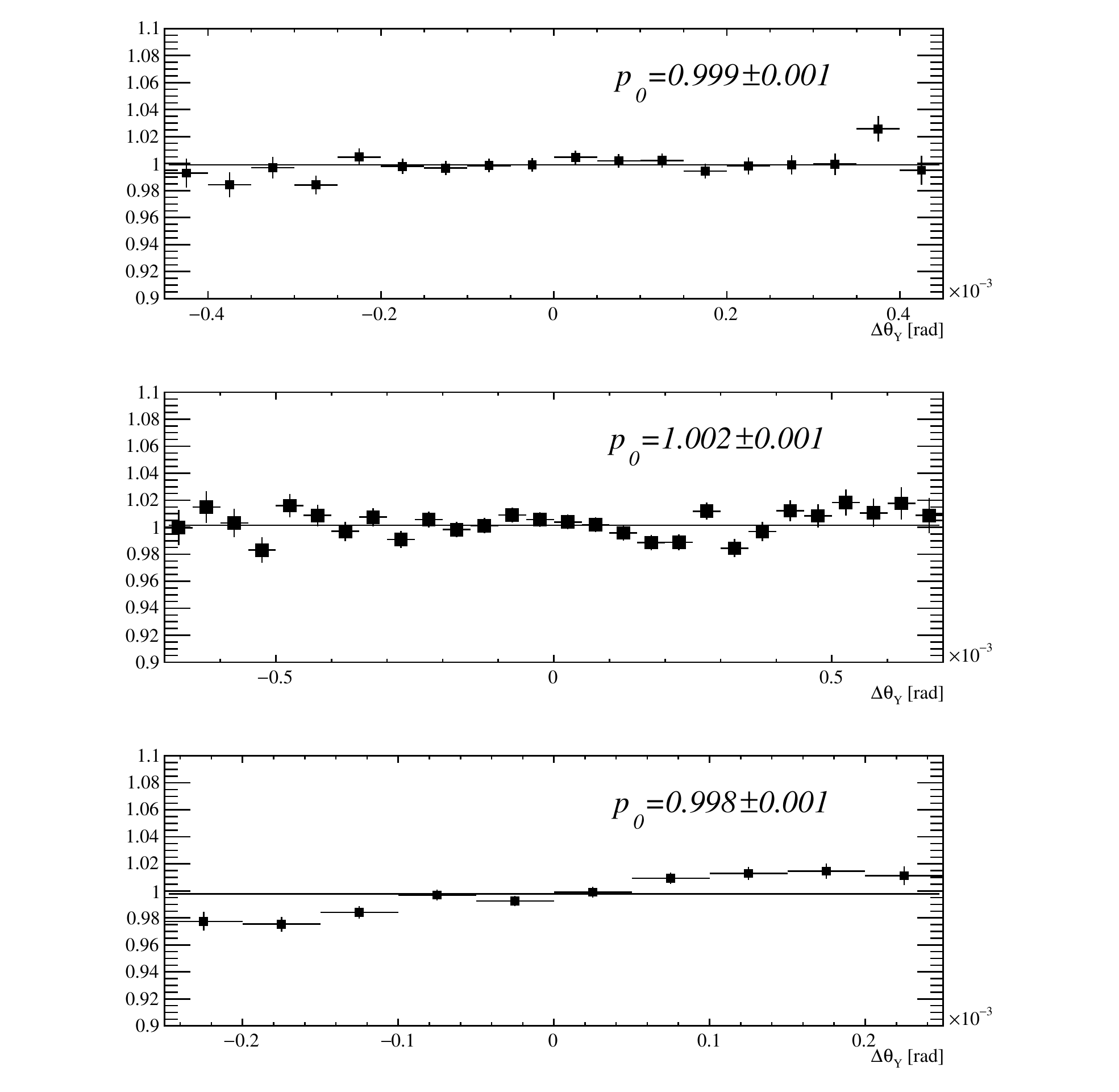}
	\caption{Data/MC ratio of $x$-projection (Left) and $y$-projection (Right) of the scattering angle between data and GEANT4 in the core region (90\% events): (upper) 12 GeV e$^-$ with 20 mm C target;
          (center) 12 GeV e$^-$ with 8 mm C target;
          (lower) 20 GeV e$^-$ with 8 mm C target.}
	\label{img:acc_cut4}
\end{figure}
Figure~\ref{img:acc_cut4}~shows the data-Monte Carlo comparison of $\Delta\theta_{X,Y}$ in the core region, which is defined to contain 90\% of the events.
A constant fit shows an agreement at the per mille level for all energies and target thickness.
For 20 GeV electrons $\Delta\theta_{Y}$ shows a 2\% asymmetry, which may be due to difference in the selection of the events between data and Monte Carlo.



\section{Multiple scattering modelling}
Modelling of the multiple scattering distribution allows a quantitative comparison with the simulation especially in the tail region where a bin-by-bin comparison suffers from the low statistics from the alignment runs.
In literature different models have been proposed for the multiple scattering~\cite{models}. The Particle Data Group (PDG)~\cite{Tanabashi:2018oc} uses a Gaussian approximation for the central 98\% of the projected angular distribution (referred to in the following as $RMS_{98}$) based on~\cite{Highland:1975pq}. 

To fit our data we will consider the model proposed in~\cite{Berger:2014fsa} which consists of a sum of a Gaussian and a Student's $t$ distribution which seems to
be suitable to describe our data.
\begin{figure}
	\centering
	\includegraphics[width=6cm]{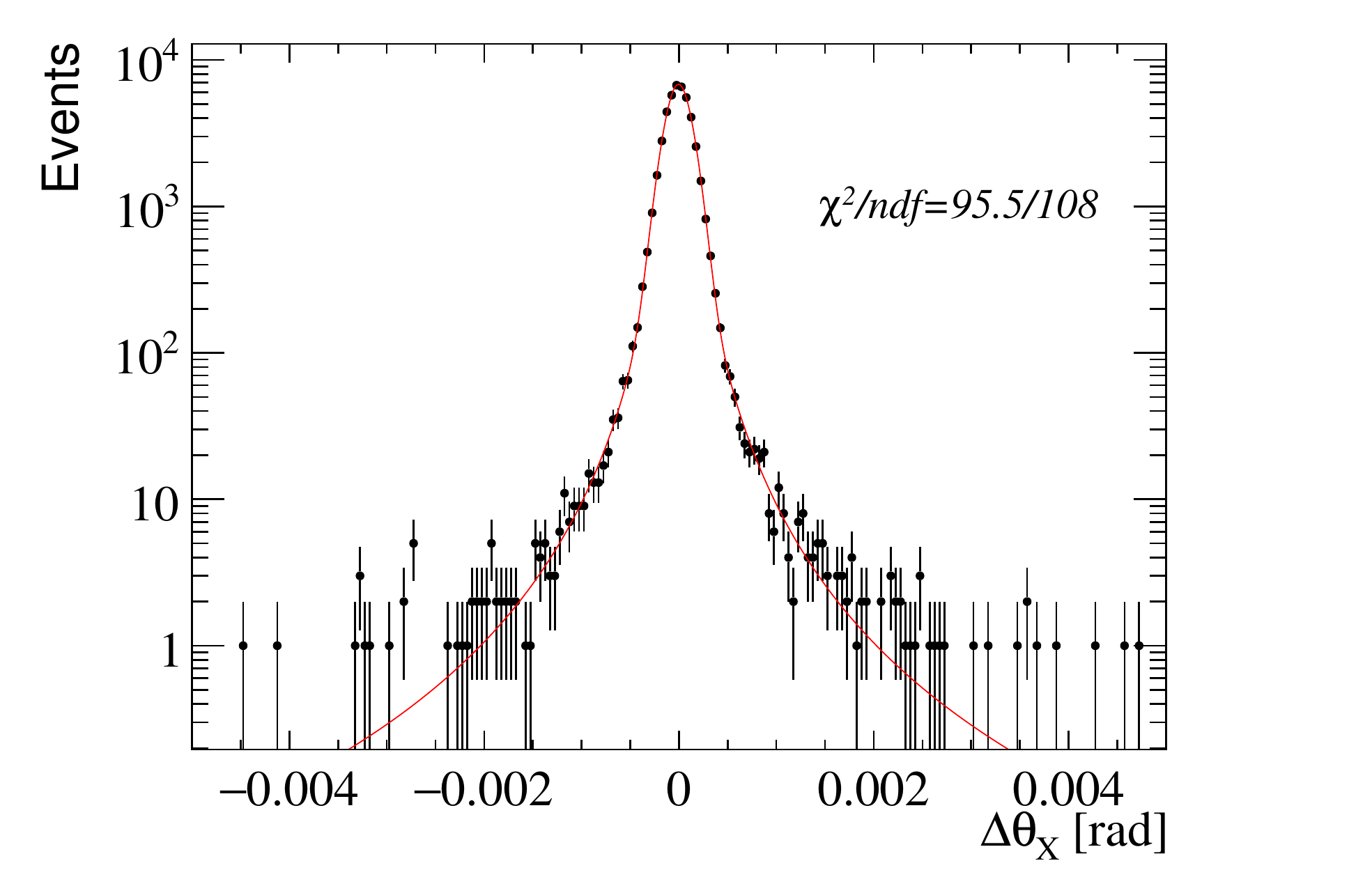}
        \includegraphics[width=6cm]{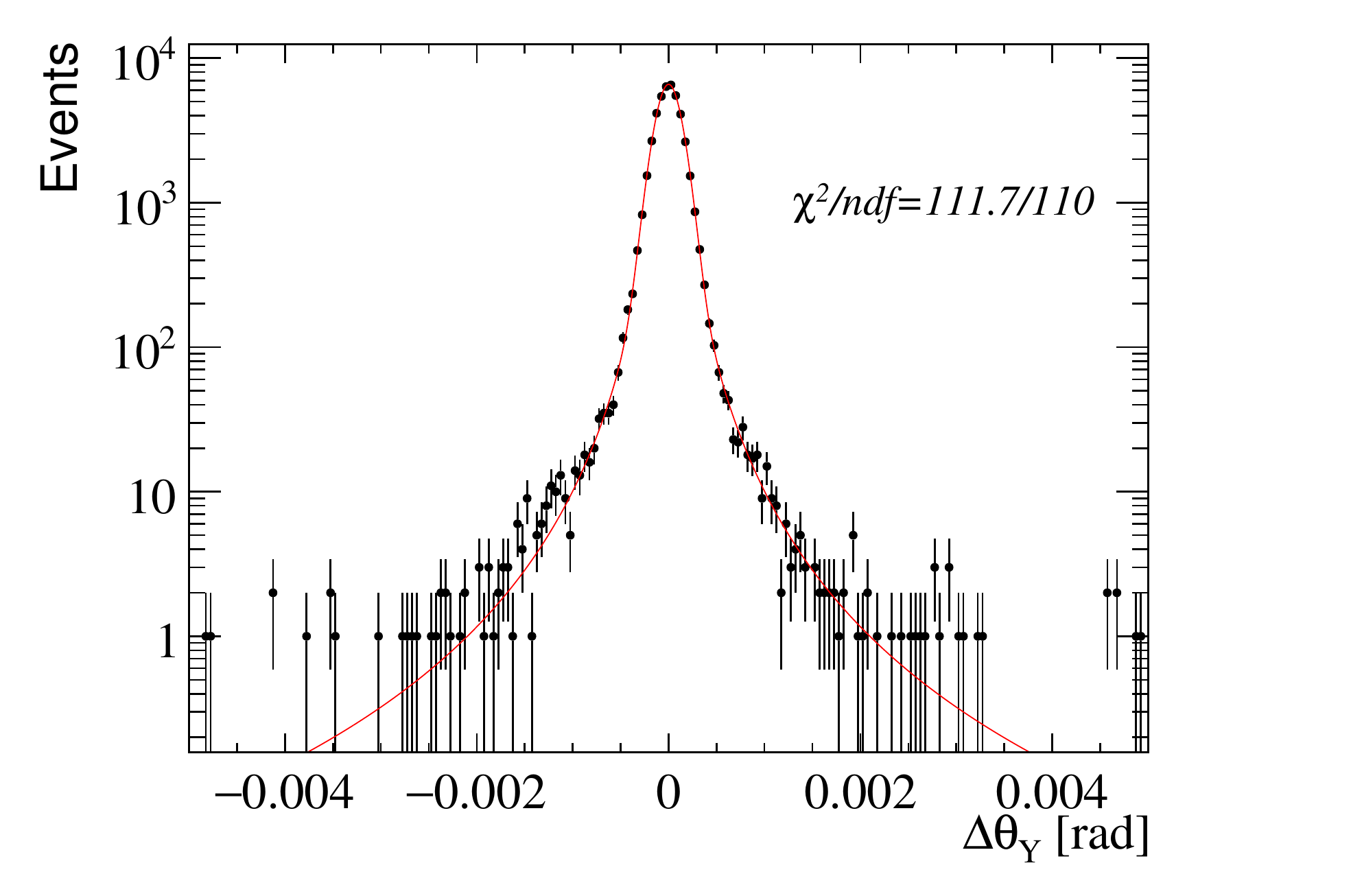}
	\caption{$x$-projection (Left) and $y$-projection (Right) of the scattering angle from 12 GeV $e^-$ alignment data ({\it i.e. without target}) compared with the results of the fit.} 
	\label{12th0}
\end{figure}
\begin{figure}
	\centering
	\includegraphics[width=6cm]{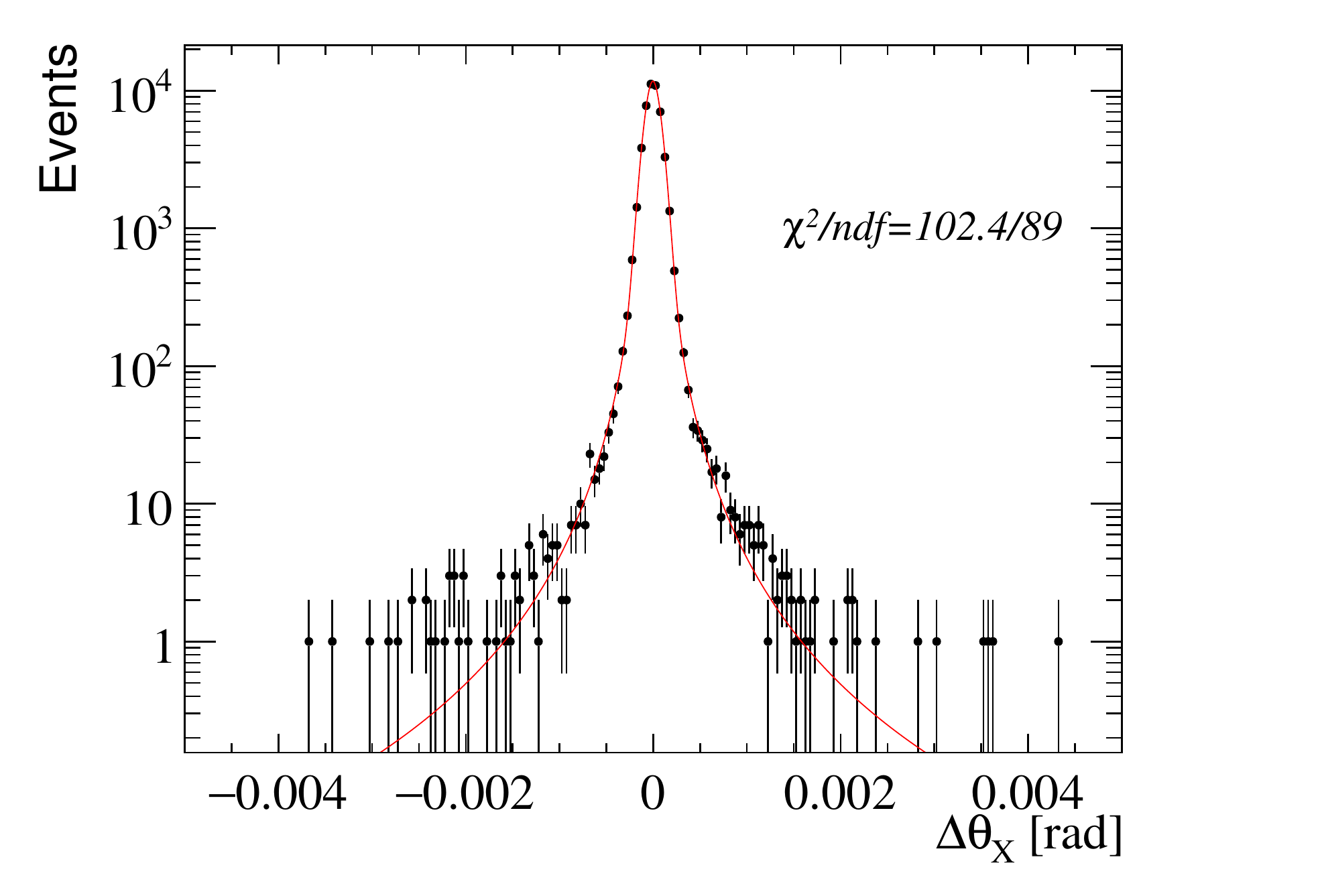}
        \includegraphics[width=6cm]{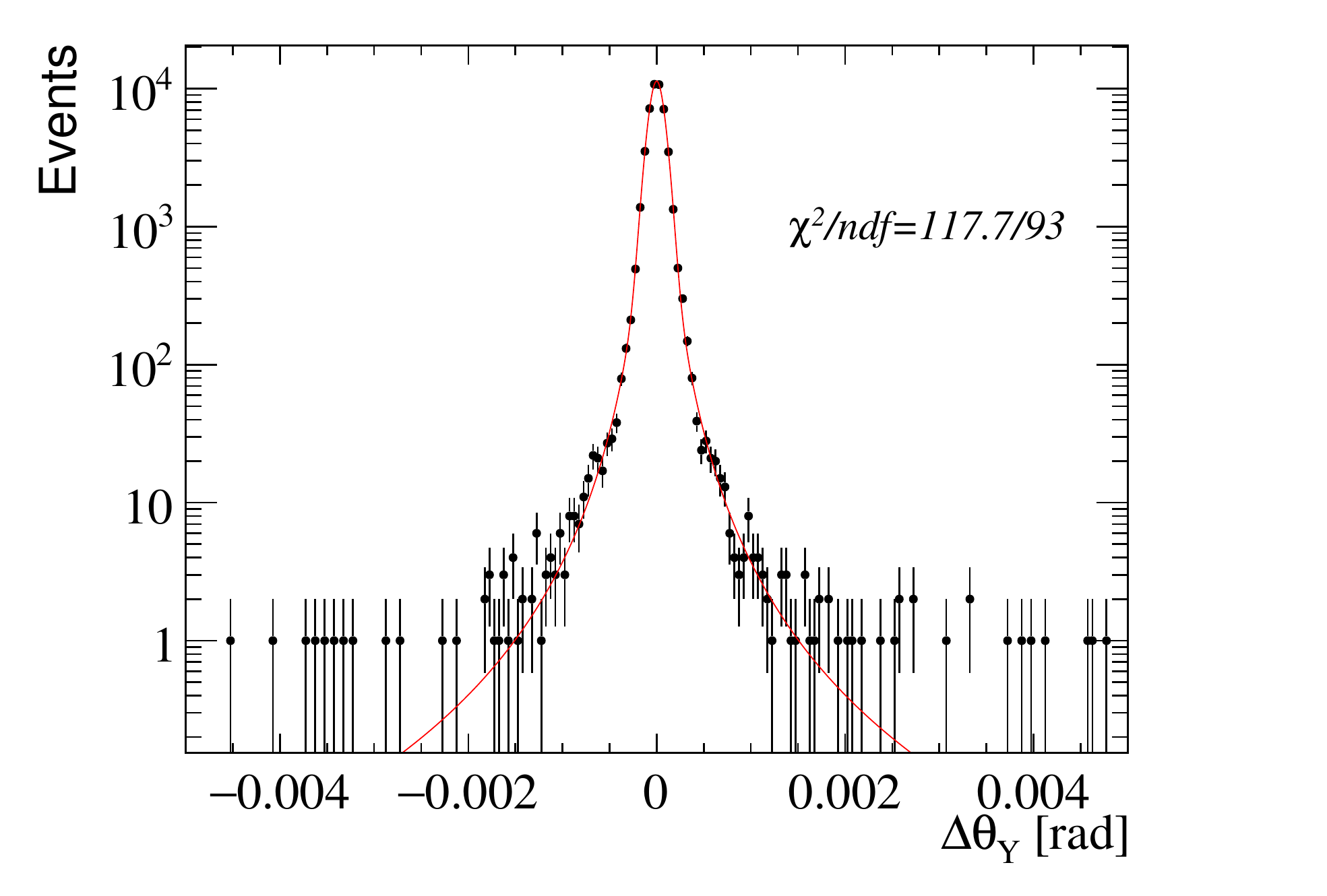}
	\caption{$x$-projection (Left) and $y$-projection (Right) of the scattering angle from 20 GeV alignment $e^-$ data ({\it i.e. without target}) compared with the results of the fit.} 
	\label{20th0}
\end{figure}


Figures~\ref{12th0},\ref{20th0}~show the result of the fit for $\Delta\theta_{X,Y}$ for runs without the target, where:
\begin{multline}
f_{\text{telescope}}(\theta) = N\left((1-a)\frac{1}{\sqrt{2\pi}\sigma_G}e^{-\frac{(\theta-\mu)^2}{2\sigma_G^2}}+a\frac{\Gamma(\frac{\nu+1}{2})}{\sqrt{\nu\pi}\sigma_T\Gamma(\frac{\nu}{2})}\left(1+\frac{(\theta-\mu)^2}{\nu\sigma_T^2}\right)^{-\frac{\nu+1}{2}}\right) .
\label{f_telescope}
\end{multline}
$ N $ is the overall normalization, $ \sigma_T$ and $ \sigma_G $ are the widths of the Student's $ t $ and Gaussian distributions respectively, $ \nu $ is the tail parameter of the $ t $ distribution, $ \mu $ the common mean and $ a $ is the relative fraction of the Student's $ t $ distribution.
As can be seen a good description is obtained  for both $\Delta\theta_{X}$ and $\Delta\theta_{Y}$ with the model. Results of the fit for alignment data are given in Table~\ref{table_align}.

\begin{table}[hp!]
\begin{center}
    \begin{tabular}{|rr|rrrr|} \hline
      $Angle$ & $E$ & $a$ & $\sigma_G$ & $\nu$ & $\sigma_T$ \\
        & GeV & & 0.1 mrad &  & 0.1 mrad \\
  \hline
  $\Delta\theta_X$ & 12 & $0.35\pm 0.02$ & $1.33\pm0.02$ & $2.2\pm0.1$ &
  $1.22\pm 0.05$ \\
\hline
  $\Delta\theta_Y$ & 12 & $0.34\pm 0.02$ & $1.32\pm0.02$ & $2.2\pm0.1$ &
$1.28\pm 0.06$ \\
   \hline
  $\Delta\theta_X$ & 20 & $0.35\pm 0.02$ & $0.84\pm0.01$ & $2.1\pm0.1$ &
   $0.73\pm 0.03$ \\
   $\Delta\theta_Y$ & 20 & $0.37\pm 0.02$ & $0.83\pm0.01$ & $2.2\pm0.1$ &
   $0.76\pm 0.03$ \\
\hline
\end{tabular}
\end{center}
\caption{Results of the fit for alignment data ({i.e. without target}). Uncertainties are statistical.}
\label{table_align}
\end{table}

Following Eq.~\ref{eq_conv}, the data with the carbon targets are fitted by the convolution of the function describing the effect of the apparatus and the target:
\begin{equation}
  f(\theta)=\int f_{\text{telescope}}(\theta-\theta') f_{\text{target}}(\theta')d\theta'
 \end{equation} 
where the function (\ref{f_telescope}) has been used both for the telescope (with parameters fixed to the alignment data (see Figs.~\ref{12th0},\ref{20th0})) and for the target, with free parameters. 

\begin{figure}
	\centering
	\includegraphics[width=6cm]{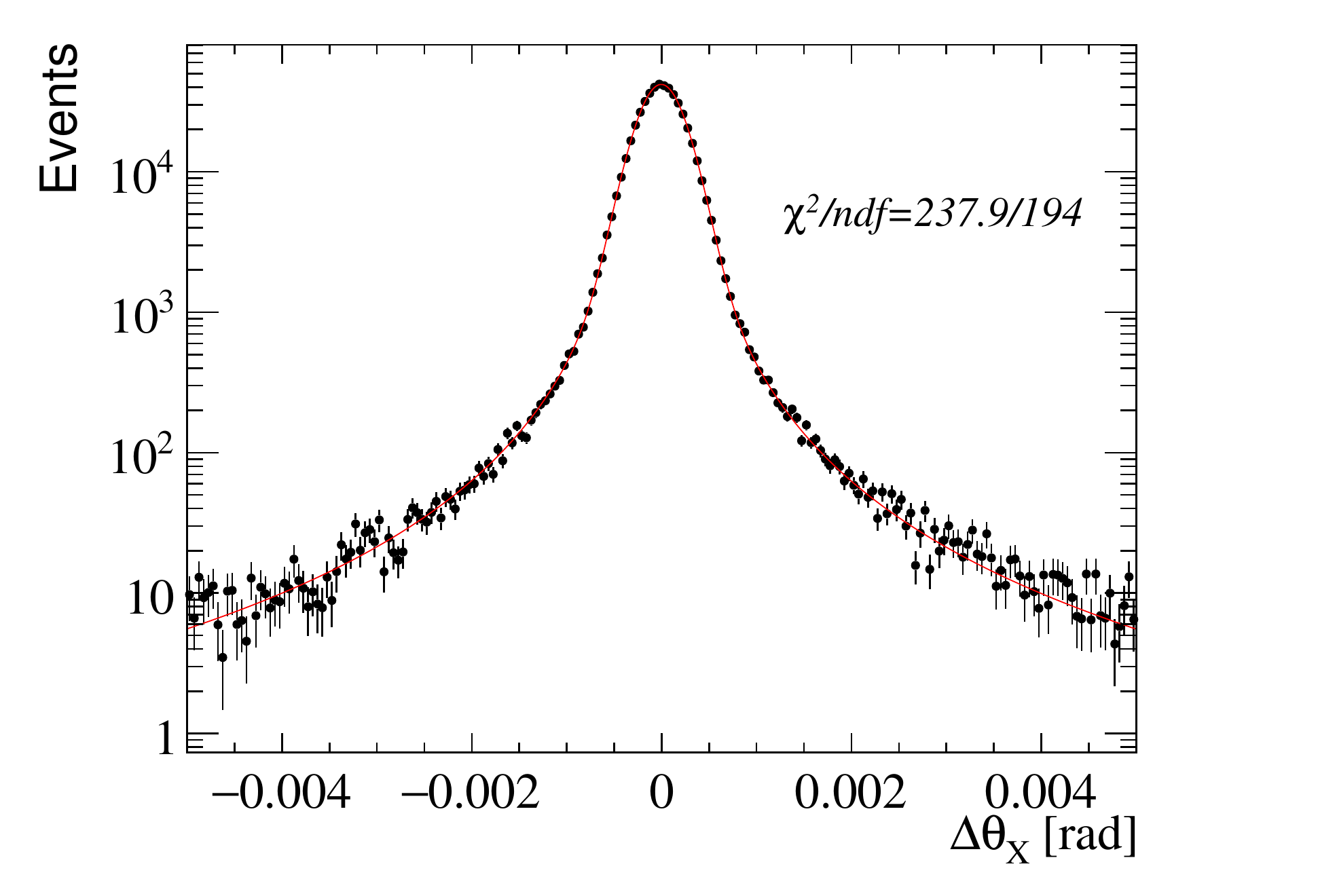}
        \includegraphics[width=6cm]{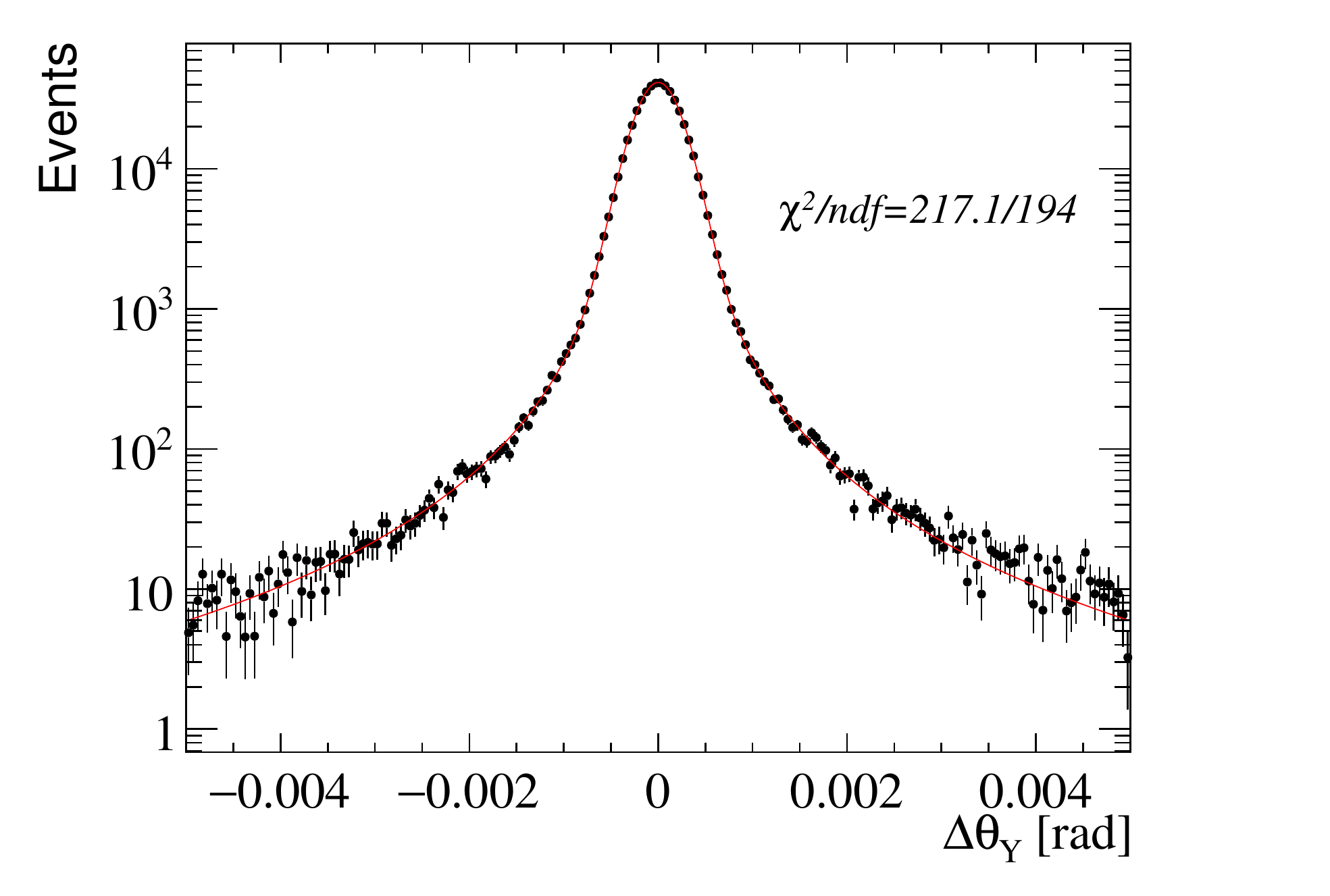}
	\caption{$x$-projection (Left) and $y$-projection (Right) of the scattering angle from 12 GeV $e^-$ for data with 8 mm target compared with the results of the fit. Details are given in the text.} 
	\label{12th8}
\end{figure}
\begin{figure}
	\centering
	\includegraphics[width=6cm]{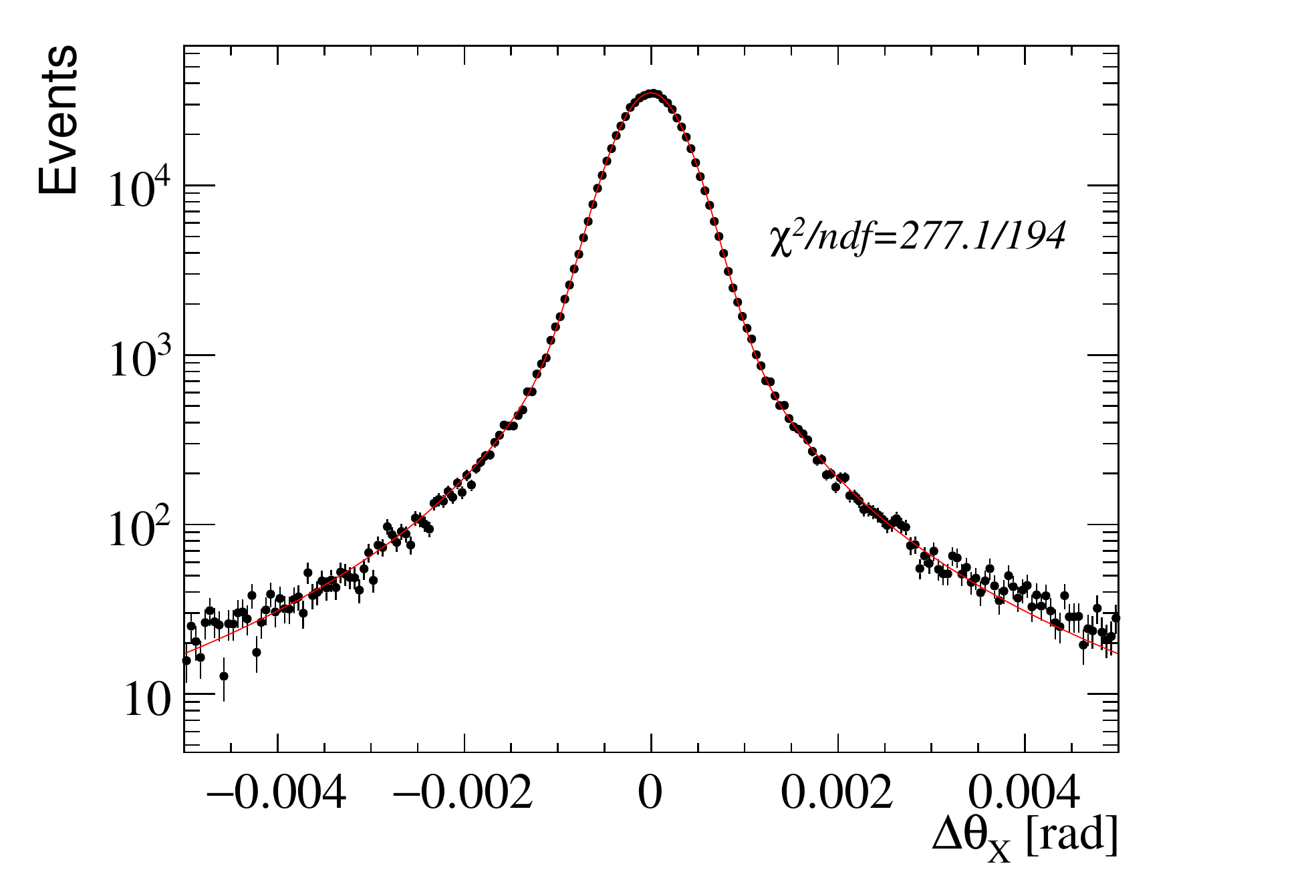}
        \includegraphics[width=6cm]{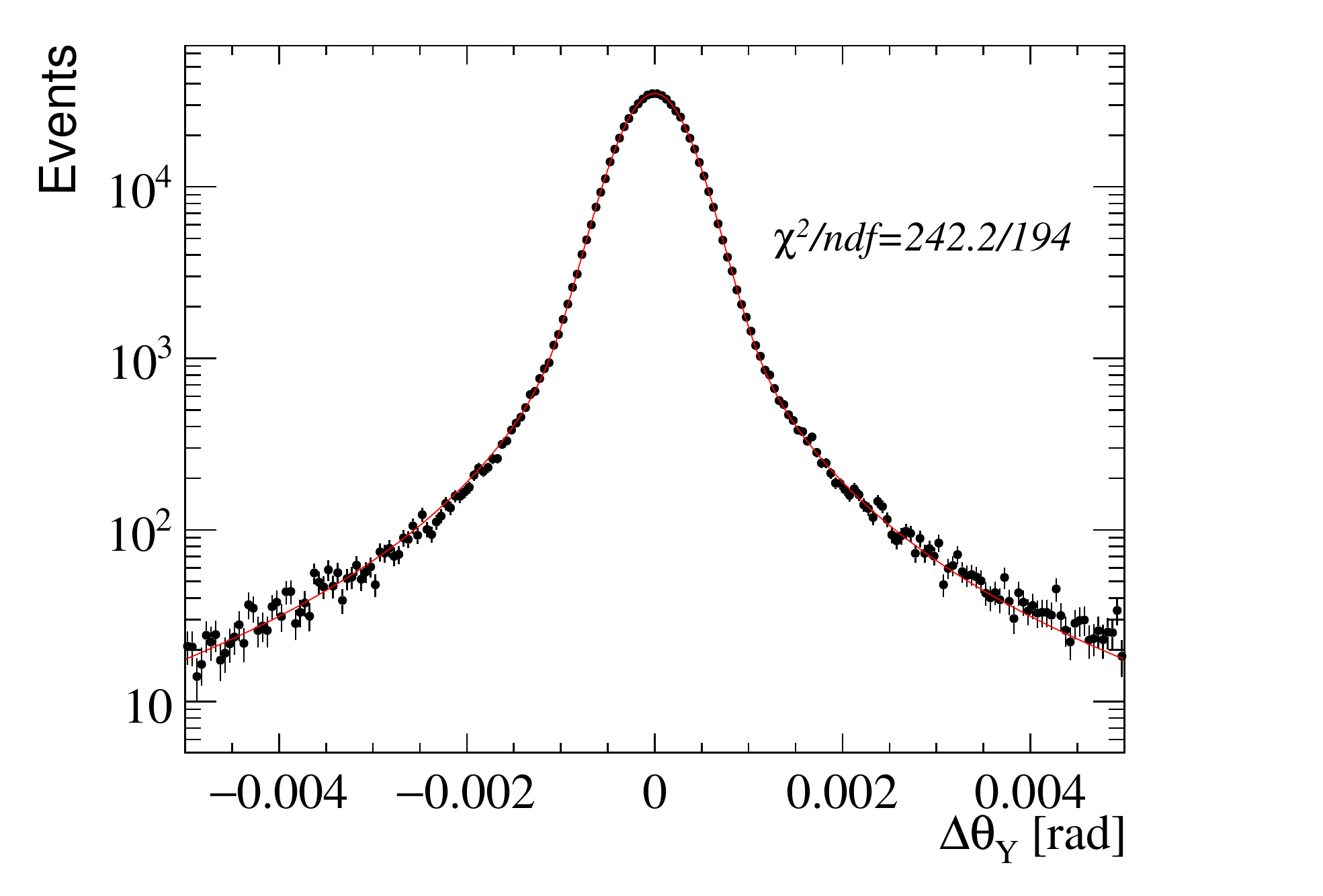}
	\caption{$x$-projection (Left) and $y$-projection (Right) of the scattering angle from 12 GeV $e^-$ for data with 20 mm target compared with the results of the fit. Details are given in the text.} 
	\label{12th20}
\end{figure}
\begin{figure}
	\centering
	\includegraphics[width=6cm]{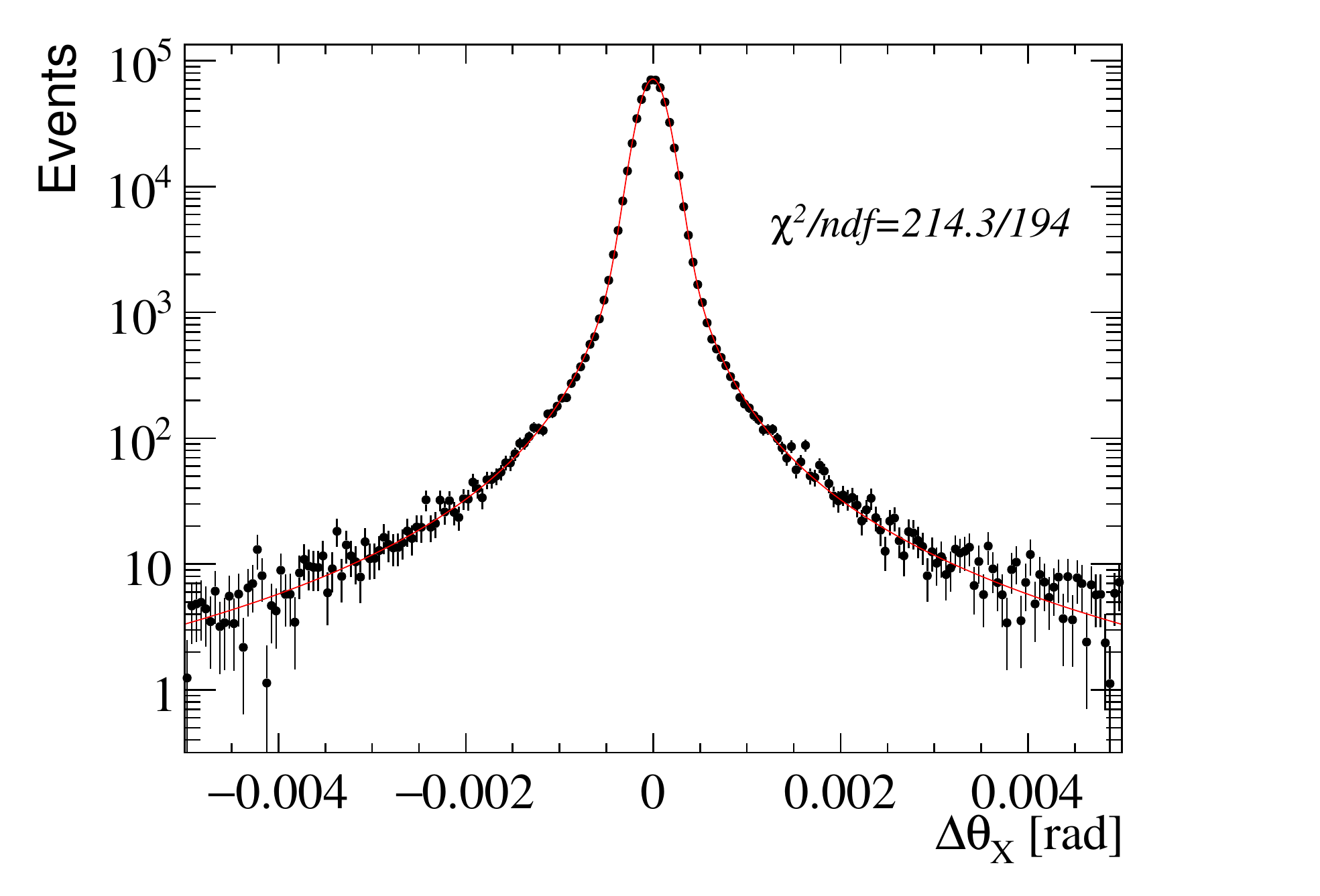}
        \includegraphics[width=6cm]{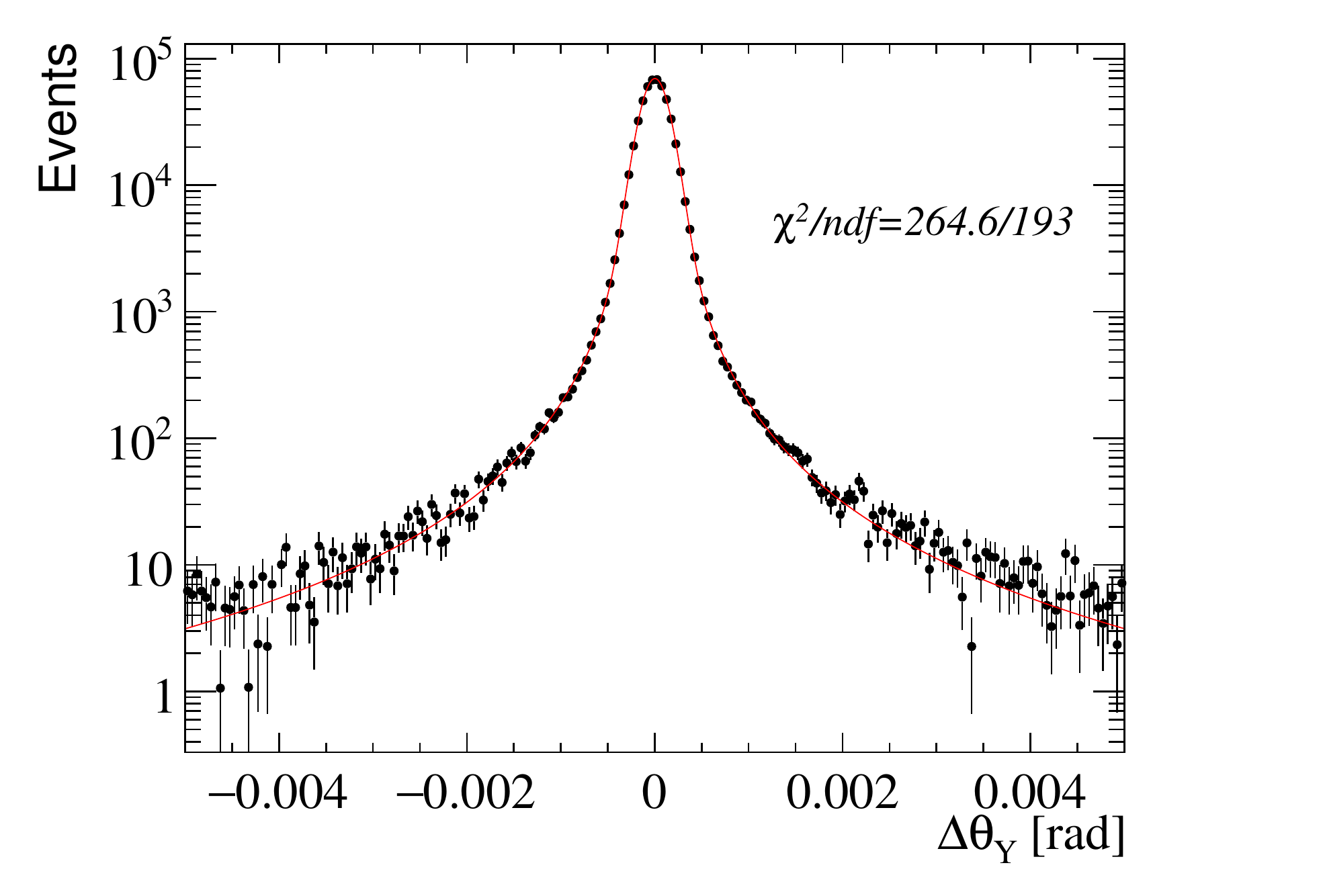}
	\caption{$x$-projection (Left) and $y$-projection (Right) of the scattering angle from 20 GeV $e^-$ for data with 8 mm target compared with the results of the fit. Details are given in the text.} 
	\label{20th8}
\end{figure}

\begin{figure}[t]
	\centering
	\includegraphics[width=6cm]{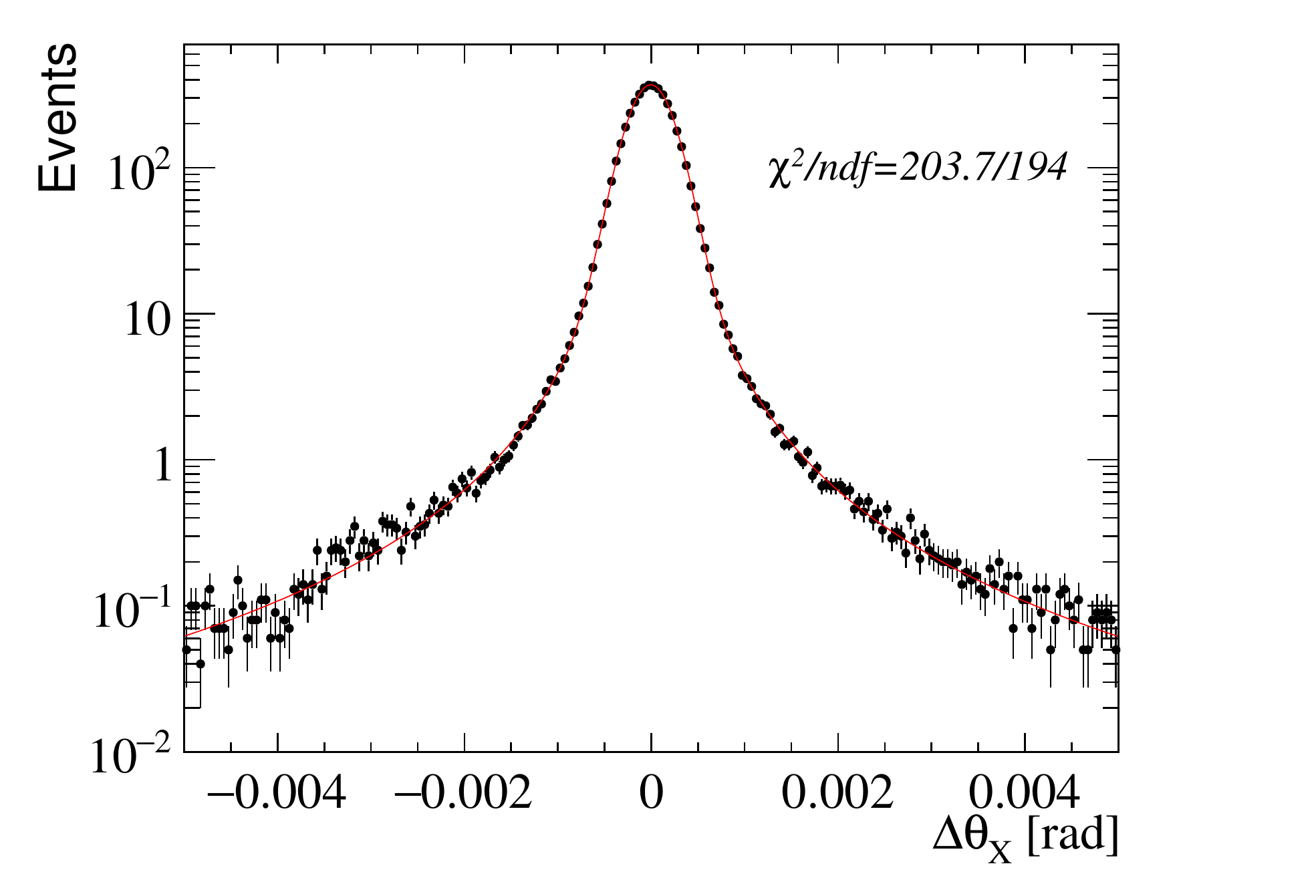}
        \includegraphics[width=6cm]{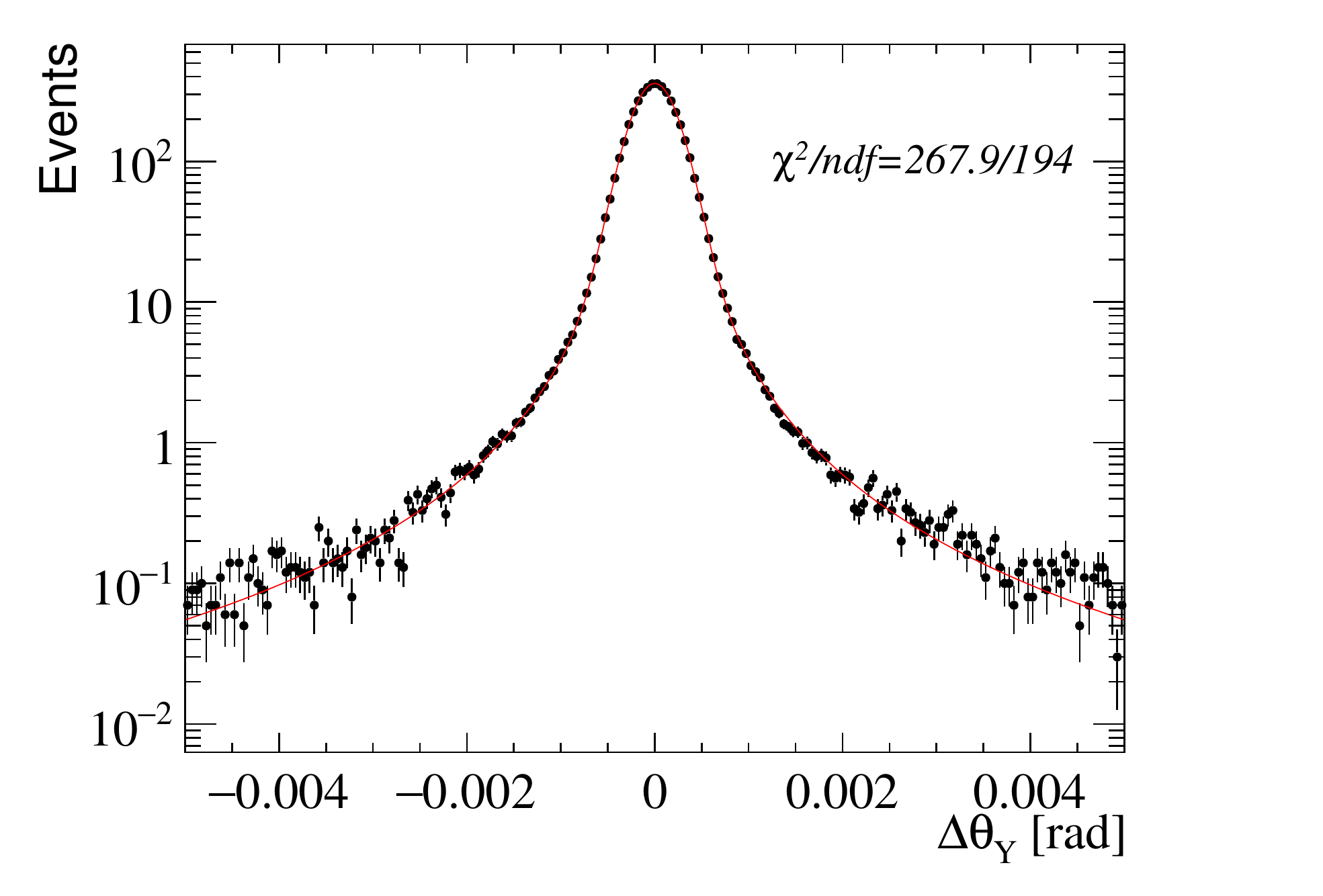}
	\caption{$x$-projection (Left) and $y$-projection (Right) of the scattering angle from 12 GeV $e^-$ from MC GEANT4 simulation with 8 mm target compared with the results of the fit. Details are given in the text.} 
	\label{12th8mc}
\end{figure}

\begin{figure}
	\centering
	\includegraphics[width=6cm]{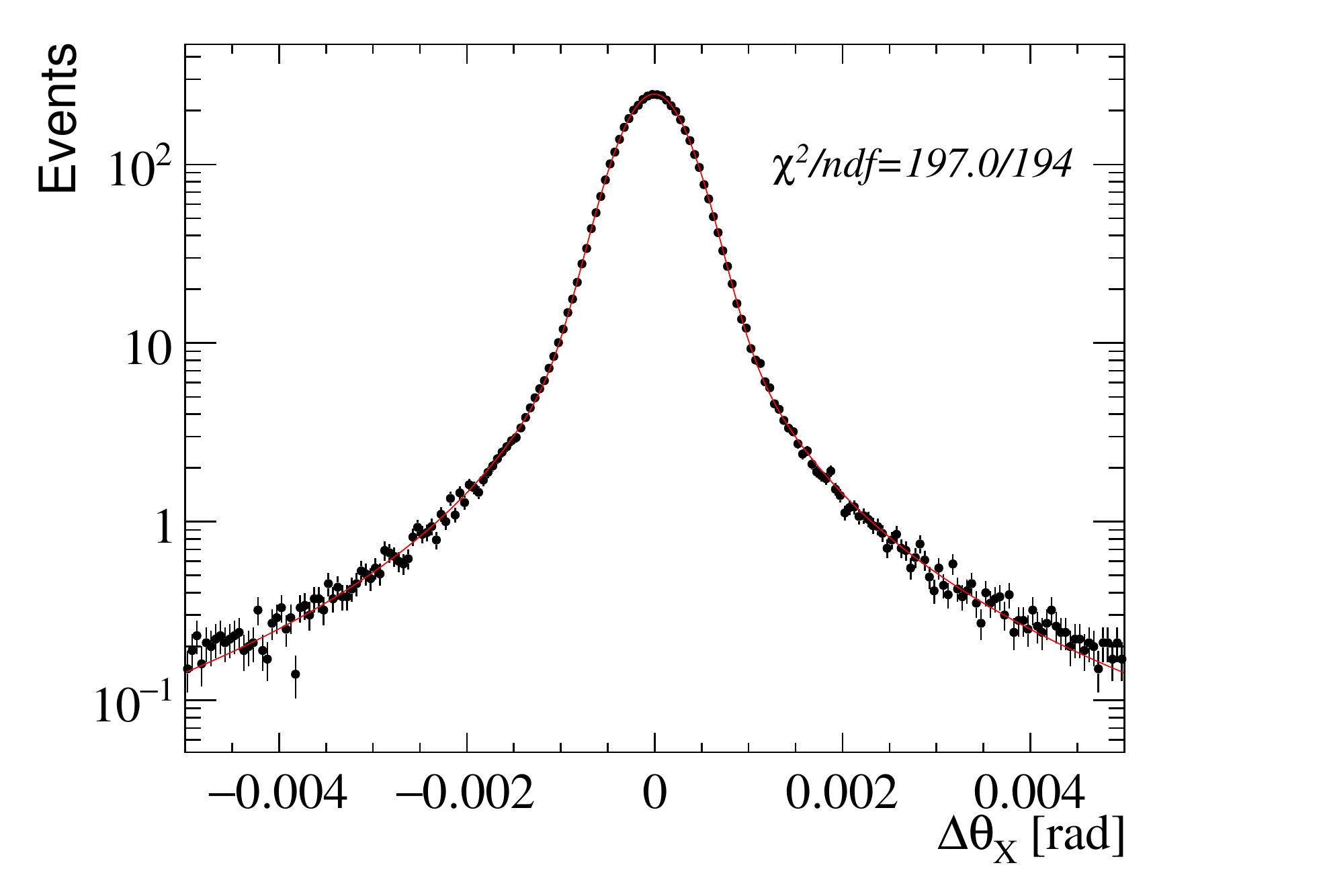}
        \includegraphics[width=6cm]{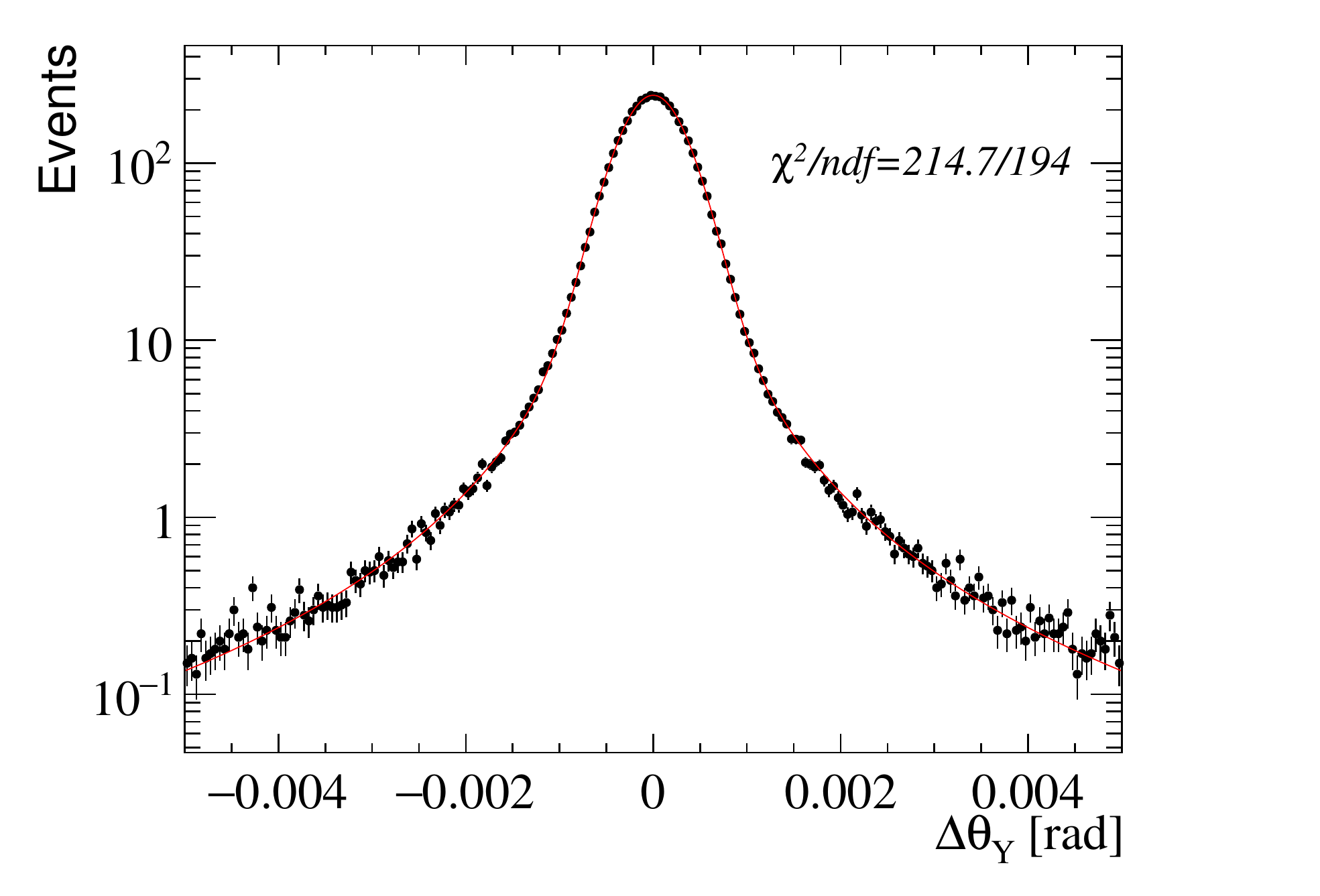}
	\caption{$x$-projection (Left) and $y$-projection (Right) of the scattering angle from 12 GeV $e^-$ from MC GEANT4 simulation with 20 mm target compared with the results of the fit. Details are given in the text.} 
	\label{12th20mc}
\end{figure}

\begin{figure}
	\centering
	\includegraphics[width=6cm]{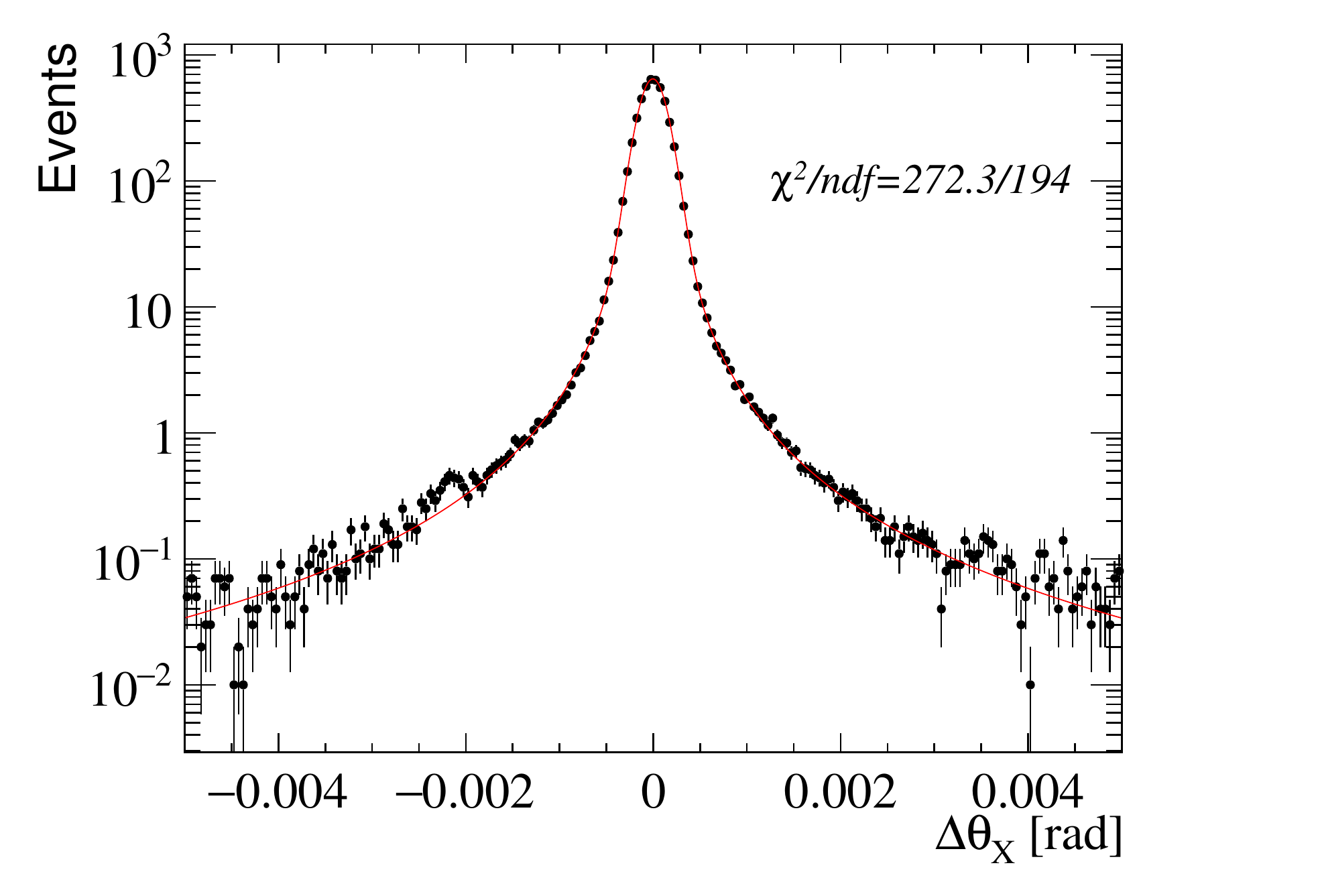}
        \includegraphics[width=6cm]{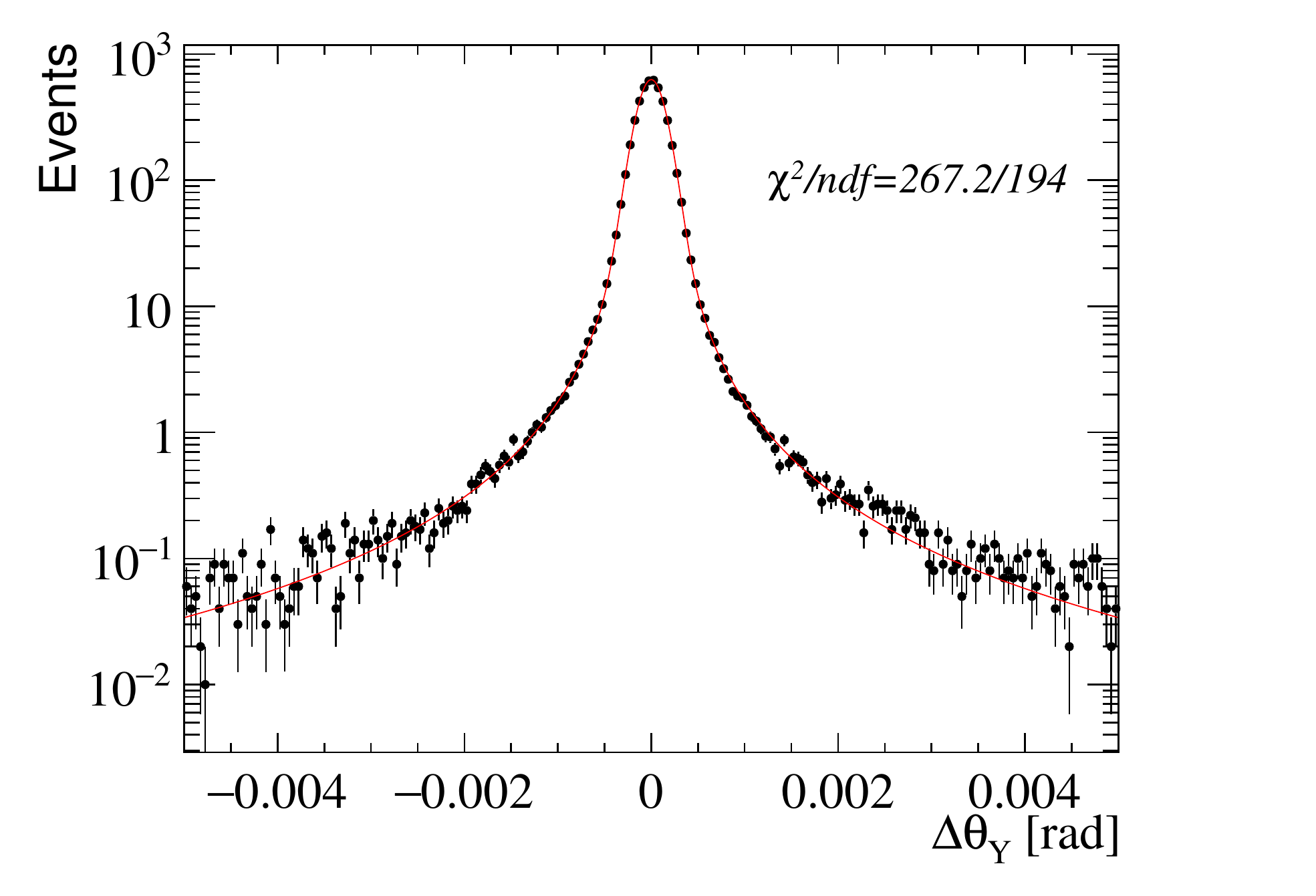}
	\caption{$x$-projection (Left) and $y$-projection (Right) of the scattering angle from 20 GeV $e^-$ from MC GEANT4 simulation with 8 mm target compared with the results of the fit. Details are given in the text.} 
	\label{20th8mc}
\end{figure}

Figures \ref{12th8}, \ref{12th20}, \ref{20th8}  show the results for the data with target, while Figs. \ref{12th8mc}, \ref{12th20mc}, \ref{20th8mc} show the results for the Monte Carlo simulation (obtained using Eq.~\ref{eq_conv}).


As can be seen, also in this case the data seem to be well represented by the fitted function. Results of the fit for data and Monte Carlo are given in Table~\ref{table_results}. The discrepancy between the parameters is below 3\% for the Gaussian width $\sigma_G$ (statistical uncertainty of ~2\%) while it increases up to 10\% on $\nu$ and to 15\% on $\sigma_T$
for the $t$ parameters, although with larger statistical errors.
$RMS_{98}$ has also been computed from the central 98\% of the scattering angle distribution with the fitted model for the target. Table~\ref{table_RMS98} shows $RMS_{98}$ obtained by the MC distribution (without any fit) of the scattering angle. The agreement between data and Monte Carlo simulation is within 2\%.

\begin{table}[hp!]
\begin{center}
 \tiny
    \begin{tabular}{|rrr|rrrr|rrrr|} \hline
  & & & \multicolumn{4}{|c|}{data} & \multicolumn{4}{|c|}{MC (GEANT 10.4)} \\
  $Angle$ & $E$ & $d$ & $a$ & $\sigma_G$ & $\nu$ & $\sigma_T$& $a$ & $\sigma_G$ & $\nu$  & $\sigma_T$\\
  & GeV & mm & & 0.1 mrad &  & 0.1 mrad & & 0.1 mrad & & 0.1 mrad \\
  \hline
  $\Delta\theta_X$ & 12 & 8 & $0.34\pm 0.01$ & $1.99\pm0.02$ & $1.52\pm0.03$ &
  $1.48\pm 0.05$ & $0.29\pm 0.01$  & $1.93\pm0.02$ & $1.39\pm 0.03$ & $1.59\pm 0.07$ \\
  $\Delta\theta_Y$ & 12 & 8 & $0.32\pm 0.01$ & $1.99\pm0.02$ & $1.45\pm0.03$ &
  $1.43\pm 0.05$ & $0.29\pm 0.01$ & $1.92\pm0.02$ & $1.46\pm 0.03$ &  $1.64\pm 0.08$ \\
  \hline
  $\Delta\theta_X$ & 12 & 20 & $0.35\pm 0.01$ & $3.21\pm0.02$ & $1.57\pm0.03$ &
  $2.95\pm 0.06$ & $0.32\pm 0.01$  & $3.11\pm0.02$ & $1.50\pm 0.03$ & $3.22\pm 0.10$  \\
  $\Delta\theta_Y$ & 12 & 20 & $0.35\pm 0.01$ & $3.21\pm0.02$ & $1.55\pm0.03$ &
  $2.92\pm 0.06$ & $0.34\pm 0.01$ & $3.17\pm0.02$ & $1.49\pm 0.03$ & $3.01\pm 0.08$   \\
  \hline
  $\Delta\theta_X$ & 20 & 8 & $0.34\pm 0.01$ & $1.20\pm0.01$ & $1.39\pm0.02$ &
  $0.82\pm 0.02$ & $0.29\pm 0.01$ & $1.16\pm0.01$ & $1.37\pm 0.02$ & $0.96\pm0.04$ \\
  $\Delta\theta_Y$ & 20 & 8 & $0.32\pm 0.01$ & $1.18\pm0.01$ & $1.42\pm0.02$ &
  $0.88\pm 0.03$ & $0.29\pm 0.01$ & $1.18\pm0.01$ & $1.31\pm 0.02$ & $0.88\pm0.04$ \\
\hline
 
  \hline
\end{tabular}
\end{center}
\caption{Comparison between data and Monte Carlo for the results of the fit (uncertainties are statistical).}
\label{table_results}
\end{table}

\begin{table}[hp!]
\begin{center}
    \begin{tabular}{|rrr|r|r|r|} \hline
  & & & data & MC (GEANT4 10.4) & MC (no Fit)\\
  $Angle$ & $E$ & $d$ & $RMS_{98}$ & $RMS_{98}$ & $RMS_{98}$\\
  & GeV & mm & 0.1 mrad & 0.1 mrad & 0.1 mrad \\
      \hline
%
      $\Delta\theta_X$ & 12 & 8  & $2.103\pm0.004$ & $2.117\pm0.004$ & $2.111\pm 0.002$\\
      $\Delta\theta_Y$ & 12 & 8  & $2.104\pm0.004$ & $2.108\pm 0.004$ & \\
      \hline
      $\Delta\theta_X$ & 12 & 20 & $3.719\pm0.005$ & $3.796\pm0.006$ &  $3.808\pm 0.003$\\
      $\Delta\theta_Y$ & 12 & 20 & $3.712\pm0.005$ & $3.769\pm0.006$ &  \\
      \hline
      $\Delta\theta_X$ & 20 & 8 & $1.262\pm0.002$ & $1.274\pm0.002$ & $1.270\pm 0.001$\\
      $\Delta\theta_Y$ & 20 & 8 & $1.268\pm0.002$ & $1.270\pm0.002$ &  \\
  \hline
\end{tabular}
\end{center}
\caption{$RMS$ of the central 98\% of the scattering angle as obtained by the fitted functions and by the MC distribution of the target.}
\label{table_RMS98}
\end{table}



\section{Discussion on systematic errors}
Different sources of systematic uncertainties affect the $\Delta\theta_{X,Y}$ data-Monte Carlo comparison:
\begin{itemize}
\item {\bf Target}: The thickness of the target has been measured with an uncertainty of about 50 $\mu$m, while an uncertainty of about 1\% has been considered for the material density;
\item {\bf Beam energy}: 
  An uncertainty of 10\% on the energy spread (assumed to be 3\%) is estimated to have an effect of about 0.2\% on $\Delta\theta_{X,Y}$ data-Monte Carlo comparison, while the beam energy scale is known with an uncertainty of 1\%~\cite{cern};
\item {\bf Detector acceptance}: A difference in the acceptance between data and Monte Carlo can account for an asymmetry of the scattering angle as is visible in 20 GeV data. We conservatively assign 50\% of this asymmetry as a systematic error due to acceptance determination;
  
\item {\bf Alignment}: An uncertainty of 1 mm on the longitudinal distance between the sensors translates into a 0.2\% systematic error;

\item {\bf Air density}: A 10\% variation of the air density due to environmental effects can lead in 10~m to a 0.3\% energy variation at target entrance;
  
\item {\bf Monte Carlo simulation}:
 A comparison with GEANT4 version \texttt{10.5}, where Mott corrections to
    $e^{\pm}$ scattering at high energy were included by default, shows no significant differences.
\end{itemize}
Table~\ref{table_sys} summarizes the systematic errors contributing to the data-Monte Carlo comparison of $\Delta\theta_{X,Y}$ in the core region (90\% of the events).
 \begin{table}[hp!]
\begin{center}
    \begin{tabular}{|c|c|c|c|} \hline
  Source & 12 GeV 8 mm & 12 GeV 20 mm & 20 GeV 8 mm\\
  \hline
  Target density & \multicolumn{3}{|c|}{0.5\%}\\ \cline{2-4}
  Target thickness & 0.3\% & 0.1\% & 0.3\% \\\cline{2-4}
  Beam energy spread &  \multicolumn{3}{|c|}{0.2\%} \\\cline{2-4}
  Beam energy scale  &  \multicolumn{3}{|c|}{1.0\%} \\\cline{2-4}
  Detector acceptance & \multicolumn{2}{|c|}{negligible} & 1\% \\\cline{2-4}
  Alignment & \multicolumn{3}{|c|}{ 0.5\%}  \\\cline{2-4}
  Air density&  \multicolumn{3}{|c|}{0.3\% } \\\cline{2-4}
  Monte Carlo &  \multicolumn{3}{|c|}{negligible} \\ \cline{2-4}
  \hline
Total  & 1.3\% & 1.3\% & 1.6\% \\
      \hline
\end{tabular}
\end{center}
\caption{List of systematic errors contributing to data-Monte Carlo comparison of $\Delta\theta_{X,Y}$ in the core region (90\% of the events).}
\label{table_sys}
\end{table}
 
 \section{Conclusion}
 The multiple scattering effects of 12 and 20 GeV electrons on 8 and 20 mm thickness carbon targets have been studied at the H8 line at CERN using the high-resolution silicon microstrip detectors of the UA9 apparatus. $x$ and $y$ projections of the scattering angle have been compared with Monte Carlo simulation and agreement below 1\%  is found on the core of the distributions (covering 90\% of the events). Differences between data and Monte Carlo reach ~10\% in the tails of the distribution of scattering angle, although with large statistical uncertainties and contributions from systematic effects due to misalignment and acceptance. Data and Monte Carlo distributions are fitted with a  model based on a convolution of a Gaussian and a Student's t functions showing agreement within 3\% (with 2\% combined statistical and systematic uncertainty) on the Gaussian width.  The comparison of the Student's t parameters which describes the tails of the distribution shows larger differences but is less precise.
 The $RMS$ of the central 98\% of the fitted distribution agrees quite well with GEANT4. These results 
 show that GEANT4 simulation meets the MUonE experimental requirements in the core region of the scattering angle distribution, while more data are needed for a quantitative comparison in the tails.

\acknowledgments
We warmly thank Marco Calviani for his help in providing us the graphite targets, Alberto Ribon
for useful discussions on GEANT, Luca Cavoto and the whole UA9 collaboration for support.



\end{document}